\documentclass[onecolumn,sort&compress,numbers]{els-mrw} 

\usepackage{amsmath,amssymb,amsfonts,amsthm,makeidx,graphicx}
\usepackage{txfonts}
\usepackage{helvet}
\usepackage{braket}
\usepackage{mathtools}
\usepackage{bbm, dsfont}
\usepackage{xspace}
\usepackage{wrapfig,lipsum}
\newcommand{\dd}{\displaystyle}
\newcommand{\nn}{\nonumber}

\newcommand{\drawsquare}[2]{\hbox{%
\rule{#2pt}{#1pt}\hskip-#2pt
\rule{#1pt}{#2pt}\hskip-#1pt
\rule[#1pt]{#1pt}{#2pt}}\rule[#1pt]{#2pt}{#2pt}\hskip-#2pt
\rule{#2pt}{#1pt}}
\newcommand{\Yfund}{\raisebox{-.5pt}{\drawsquare{6.5}{0.4}}}
\newcommand{\Yasymm}{\raisebox{-3.5pt}{\drawsquare{6.5}{0.4}}\hskip-6.9pt%
        \raisebox{3pt}{\drawsquare{6.5}{0.4}}}

\newcommand{\rep}[1]{\ensuremath\boldsymbol{#1}}
\newcommand{\crep}[1]{\ensuremath\overline{\boldsymbol{#1}}}
\newcommand{\Z}[1]{\ensuremath{\mathds{Z}_{#1}}} 
\newcommand{\SO}[1]{\ensuremath{\mathrm{SO}(#1)}}
\newcommand{\SU}[1]{\ensuremath{\mathrm{SU}(#1)}}

\newcommand{\U}[1]{\ensuremath{\mathrm{U}(#1)}}
\newcommand{\E}[1]{\ensuremath{\mathrm{E}_{#1}}}
\newcommand{\e}{\mathrm{e}}
\newcommand{\I}{\mathrm{i}}

\newcommand{\CP}{\ensuremath{\mathcal{CP}}\xspace}
\newcommand{\x}{\ensuremath{\times}}
\newcommand{\wavefunction}[1][j,M]{\ensuremath{\psi^{#1}}}
\newcommand{\orb}[2]{\hspace*{-0.1ex}\begin{bsmallmatrix}\!#1\!\\[0.3ex]\!#2\!\end{bsmallmatrix}\hspace*{-0.1ex}}

\DeclareMathOperator{\im}{Im}

\usepackage[pdftex,hidelinks,colorlinks=true,citecolor=darkgreen]{hyperref}
\hypersetup{
    pdftitle = {Quark and lepton masses},
    pdfauthor = {Ferruccio, Ramos-Sanchez}
}

\begin{document}


\chapter{Quark and lepton masses}\label{chap1}

\author[1]{Ferruccio Feruglio}%
\author[2]{Sa\'ul Ramos-S\'anchez}%


\address[1]{\orgname{INFN}, \orgdiv{Sezione di Padova}, \orgaddress{Via Marzolo~8, I-35131 Padua, Italy}}
\address[2]{\orgname{Instituto de F\'isica}, \orgdiv{Universidad Nacional Aut\'onoma de M\'exico}, \orgaddress{Cd. de M\'exico C.P. 04510, M\'exico}}


\maketitle

\begin{abstract}[Abstract]
	Quarks and leptons, the fundamental building blocks of the subatomic world, manifest in three families - replicas with identical quantum numbers that differ only in their masses. After summarizing the present data, an overview is presented of the main attempts to explain the origin of the observed patterns and trace it back to an as-yet-unknown fundamental principle.
\end{abstract}

\begin{keywords}
fermion masses\sep mixing angles\sep composite fermions\sep extra dimensions\sep flavor symmetry\sep string theory
\end{keywords}


\begin{glossary}[Nomenclature]
	\begin{tabular}{@{}lp{34pc}@{}}
		AdS & Anti-de Sitter\\
		CFT & Conformal Field Theory\\
		CKM & Cabibbo Kobayashi Maskawa\\
		FCNC & Flavor Changing Neutral Current\\
		FN & Froggatt Nielsen\\
		GIM & Glashow Iliopoulos Maiani\\
		GUT & Grand Unified Theory\\
		MSSM & Minimal Supersymmetric Standard Model\\
		PMNS & Pontecorvo Maki Nakagawa Sakata\\
		QCD & Quantum Chromo Dynamics\\
		RGE & Renormalization Group Equation\\
		RS & Randall Sundrum\\
		SM & Standard Model\\
		SUSY & Supersymmetry or Supersymmetric\\
		VEV & Vacuum Expectation Value
	\end{tabular}
\end{glossary}

\section*{Objectives}
\begin{itemize}
\item Review the structure of the Standard Model: its symmetries, particle content, and the generational pattern that accommodates quarks and leptons.
\item Review the current experimental data on fermion masses, mixing angles, and \CP-violating phases.
\item Motivate the search for an organizing principle beyond these measured parameters.
\item Explain why grand unified theories can provide relations between quark and lepton mass parameters.
\item Explain how the hierarchical structure of fermion masses can be reproduced through quantum corrections.
\item Illustrate how fermion compositeness can change our perspective on the fermion mass puzzle.
\item Explain how symmetries could play a role in deciphering the fermion puzzle.
\item Explain how, in string theories, fermion mass and mixing properties arise and are in principle calculable.
\end{itemize}

\section{Introduction}\label{intro}
Quarks and leptons represent the deepest layer of the microscopic world explored so far. 
This layer has revealed an unexpected structure: three generations or families or flavors that share identical interaction properties and differ only in their masses. The generations can communicate through a well-defined mixing system that carefully governs transitions among them. In the SM of fundamental interactions, a simple mechanism - stemming from interactions with the Higgs doublet - provides masses and mixing angles for quarks and leptons, while also introducing phases that distinguish the behavior of particles from that of antiparticles.
This framework is in remarkable agreement with a wide range of experimental data, yet it requires a large number of parameters. The gauge principle underlying the SM allows all interactions between fermions and gauge bosons to be described with only three coupling constants, plus two additional parameters to characterize the Higgs sector. However, gauge symmetry alone - together with Lorentz invariance - places few constraints on the Yukawa sector responsible for fermion masses. In practice, each mass, mixing angle, and phase requires an independent free parameter, for a total of 22 variables that, in principle, could take on arbitrary values. Even the number of observed generations could vary within a broad range without qualitatively altering the theory's behavior. 

Yet, most of these variables do not appear to follow a random pattern.  
The ratios of charged fermion masses within a single generation vary by one to two orders of magnitude, while the entire mass spectrum of the three families covers more than five orders of magnitude. Quark mixing angles are small and dominated by mixing between the first two generations. Neutrinos - long believed to be massless until the discovery of oscillations - have extremely small masses, at least six orders of magnitude smaller than the electron mass. In contrast, two of the three mixing angles in the lepton sector are large.
This overall picture naturally raises a fundamental question:  
Is there an underlying principle behind the structure of fermion masses and mixing angles, or is the SM already the most efficient description possible? Can these quantities be derived from a smaller set of parameters, or does Nature allow complete freedom, making it possible, in principle, to conceive a world where fermion masses are entirely different? While the second possibility cannot be excluded on logical grounds, this Chapter reviews attempts to identify a principle that could quantitatively account for the observed masses and mixings.

Naturally, this pursuit requires going beyond the SM.  
Over the past decades, several directions have been explored, which will be summarized in the various sections of this Chapter where, not attempting to cover the vast body of literature on the subject, we will focus on key ideas, illustrated through selected examples.  
Mass relations between quarks and leptons are expected in GUTs, covered in Section \ref{sec4}. Mechanisms where fermion masses arise through successive radiative corrections are illustrated in Section \ref{sec5}. The possibility that quarks and leptons are not elementary particles but composite states whose dynamics explains the observed data is reviewed in Section \ref{sec6}. Models based on symmetries acting in generation space are discussed in Section \ref{sec7}. Finally, Section \ref{sec8} considers how fermion masses could, in principle, be determined within a fundamental theory that unifies electroweak and strong interactions with gravity, such as String Theory.
As we shall see, this fascinating puzzle - one of Nature's deepest mysteries - remains unsolved.  
Many proposals contain important insights and innovative ideas that may ultimately guide us toward a solution. However, it is not yet clear which path among the many is the most promising. 

\section{Fermion masses and mixing angles}\label{sec2}
Quarks and leptons are organized into three fermion generations having the same quantum numbers. Leptons have either electric charge
$-1$: $e_i=(e,\mu,\tau)$; or are electrically neutral: $\nu_i=(\nu_e,\nu_\mu,\nu_\tau)$. Quarks have fractional electric charge: $+2/3$ for $u_i=(u,c,t)$ and $-1/3$ for $d_i=(d,s,b)$. Quantum numbers and mass define the flavor of a particle. Fermion masses are intimately related to mixing angles,
which describe flavor-changing transitions, and to phases responsible for \CP violation. The tight link among these
quantities can be appreciated recalling their origin in the SM. 
\subsection{The Standard Model}\label{sec2:subsec1}
Strong and electroweak interactions are described by the SM, a renormalizable gauge theory
invariant under local transformation of the group $G_{SM}=\SU3\x\SU2\x\U1$, with gauge coupling constants $g_S$, $g$ and $g'$.
The field content includes a spin zero Higgs doublet $\varphi$, and three generations of quarks and leptons, whose representations under the group $G_{SM}$ are listed in Table~\ref{tab:SM}.
\begin{table}[h!]
	\TBL{\caption{\label{tab:SM}Irreducible representations of $G_{SM}$ for quarks, leptons and Higgs doublet. Quarks and leptons occurs in three replica, or generations, with identical quantum numbers: ($u$, $d$, $\nu_e$, $e$) for the first generation,
($c$, $s$, $\nu_\mu$, $\mu$) for the second one and ($t$, $b$, $\nu_\tau$, $\tau$) for the third one. Fermions with the same electric charge will also be designated by a generation index $(i=1,2,3)$. For example $u_1=u$, $u_2=c$ and $u_3=t$.}}{%
		\begin{tabular*}{\columnwidth}{@{\extracolsep\fill}cccccccc@{}}
			\toprule
			\TCH{Group}&
			\multicolumn{1}{c}{$q_L=\left(\begin{array}{c} u_L\\d_L\end{array}\right)$} &
			\multicolumn{1}{c}{$u_R$} &
		        \multicolumn{1}{c}{$d_R$} &
	                \multicolumn{1}{c}{$l_L=\left(\begin{array}{c} \nu_L\\e_L\end{array}\right)$} &
			\multicolumn{1}{c}{$e_R$}	&    
			\multicolumn{1}{c}{$\varphi$}&
			\multicolumn{1}{c}{$\tilde\varphi\equiv i\sigma_2 \varphi^*$}	\\  
			\colrule
			\SU3 & $\rep3$& $\rep3$ & $\rep3$ & $\rep1$ & $\rep1$ & $\rep1$ & $\rep1$\\
			\SU2 & $\rep2$& $\rep1$ & $\rep1$ & $\rep2$ & $\rep1$ & $\rep2$ & $\rep2$\\
			\U1 & $+1/6$& $+2/3$ & $-1/3$ & $-1/2$ & $-1$ & $+1/2$&$-1/2$\\
			\botrule
	\end{tabular*}}{}
\end{table}

\noindent
The SM Lagrangian reads 
\begin{align}
{\cal L}_{SM}={\cal L}_{gauge}+{\cal L}_{\varphi}+i\sum_f \bar f \gamma^\mu D_\mu f+{\cal L}_{Y}\,,
\end{align}
where ${\cal L}_{gauge}$ describes the propagation and the self-interactions of the gauge vector bosons:
the eight gluons of strong interactions, together with the four electroweak
gauge vector bosons.
The term ${\cal L}_{\varphi}$ describes the propagation of the Higgs doublet, its interaction with the electroweak gauge bosons and its self-interaction. This part of the theory determines the vacuum state and is responsible for the spontaneous breaking of the gauge group $G_{SM}$ down to $\SU3\x\U1_{em}$. Out of the four gauge vector bosons of the electroweak sector, only the photon
$A_\mu$ remains massless, while two electrically charged vector bosons $W^\pm_\mu$ and one neutral gauge boson $Z_\mu$ become massive. 
In the third term, the sum runs over the five irreducible fermion representations ($f=q_L,u_R,d_R,l_L,e_R$) of Table~\ref{tab:SM}
and over the three generations.
Gauge invariance requires covariant derivatives $D_\mu f$. They describe the interactions between fermions and gauge vector bosons:
\begin{align}
i\sum_f \bar f \gamma^\mu D_\mu f = i\sum_f \bar f \gamma^\mu \partial_\mu f - \frac{g}{\sqrt{2}}~ (W^+_\mu J^{-\mu}+W^-_\mu J^{+\mu}) - \sqrt{g^2+{g'}^2}~ Z_\mu J_Z^\mu-e~ A_\mu J_{em}^\mu\,,
\end{align}
where the currents $J^{\pm \mu}$, $J_Z^{\mu}$ and $J_{em}^{\mu}$ read
\begin{align}
J^{-\mu}=&~\bar u_L \gamma^\mu d_L+ \bar \nu_L \gamma^\mu e_L\,,~~~~~~~~~~~~~~~~~J^{+\mu}=(J^{-\mu})^\dagger\,,&~
J_Z^\mu=&~J_{3L}^\mu-\frac{{g'}^2}{g^2+{g'}^2} J_{em}^\mu\,,\\
J_{em}^\mu=&~+\frac23 \bar u\gamma^\mu u-\frac13 \bar d\gamma^\mu d- \bar e\gamma^\mu e\,,&~
J_{3L}^\mu=&~\frac12\bar u_L \gamma^\mu u_L-\frac12\bar d_L \gamma^\mu d_L+\frac12\bar \nu_L \gamma^\mu \nu_L-\frac12\bar e_L \gamma^\mu e_L
\,.
\end{align}
A sum over the three generations, whose gauge interactions are identical, is 
understood in these expressions.
Finally, ${\cal L}_{Y}$ describes the Yukawa interactions:
\begin{align}
{\cal L}_{Y}=-\bar u_{Ri} Y^u_{ij}~ \varphi~ q_{Lj}-\bar d_{Ri} Y^d_{ij}~ \tilde\varphi~ q_{Lj}-\bar e_{Ri} Y^e_{ij}~ \tilde\varphi~ l_{Lj}+h.c. \,,
\end{align}
where the indices $i$ and $j$ run over the three generations and gauge indices are understood.
The SM Lagrangian includes all gauge-invariant terms with mass dimension smaller or equal to four.
Higher-dimensional gauge-invariant terms are excluded by the renormalizability requirement,
which can be relaxed in a more general approach. The renormalizable
SM Lagrangian enjoys an accidental global symmetry under transformations of the group $\U1_\mathrm{B}\x\U1_{\mathrm{L}_e}\x\U1_{\mathrm{L}_\mu}\x\U1_{\mathrm{L}_\tau}$, which leads to the conservation of the baryon number B and the three family lepton numbers $\mathrm{L}_e$, $\mathrm{L}_\mu$ and $\mathrm{L}_\tau$, at the classical level. The baryon number is a good approximate symmetry in Nature, while each individual lepton number, with the exception of their sum, is known to be violated in neutrino oscillations.

The quantities $Y^{u,d,e}$ are complex, $3\x3$ matrices including, a priori, 54 real parameters:
27 absolute values and 27 phases. 
Most of these parameters are unphysical, as can be seen by performing a general change of basis
in generation space, which leads to an equivalent Lagrangian in terms of new Yukawa couplings. This change of basis consists of a $\U3^5$ transformation, that is one \U3  per each irreducible representation, described by a total of 45 real parameters, 15 angles and 30 phases. The subgroup of accidental symmetries $\U1_\mathrm{B}\x\U1_{\mathrm{L}_e}\x\U1_{\mathrm{L}_\mu}\x\U1_{\mathrm{L}_\tau}$ leaves
the Lagrangian invariant, and the change of basis which can affect $Y^{u,d,e}$ is effectively
described by 15 angles and 26 phases. A corresponding number of parameters can be removed from the initial matrices $Y^{u,d,e}$,
leaving 12 absolute values and 1 phase. Their physical meaning becomes transparent
in the unitary gauge, where the Higgs doublet is parametrized as
\begin{align}
\varphi=
\left(
\begin{array}{c}
0\\
\dd\frac{h+v}{\sqrt{2}}
\end{array}
\right)\,.
\end{align}
Here $v\approx 246$ GeV is the VEV of the Higgs field, the only dimensionful parameter of the theory, hence
providing the mass scale for all the particles of the SM.
In the new basis the matrices $Y^{u,d,e}$ are diagonal, real and positive definite, and ${\cal L}_{Y}$ becomes
\begin{align}
{\cal L}_{Y}=-\left(1+\frac{h}{v}\right)\left(\bar u_{Ri}~ m^u_i~ u_{Li}+\bar d_{Ri}~ m^d_i~  d_{Li}+\bar e_{Ri}~ m^e_i~  e_{Li}\right)+h.c. \,,
\end{align}
where $m^{u,d,e}_i= Y^{u,d,e}_{ii} \frac{v}{\sqrt{2}}$ are the 9 fermion masses. 
At the same time, the change of basis modifies the charged currents $J^\pm_\mu$, which now read
\begin{align}
J^{-\mu}=&~\bar u_L \gamma^\mu V_{\mathrm{CKM}} d_L+ \bar \nu_L \gamma^\mu e_L\,,\qquad
J^{+\mu}=(J^{-\mu})^\dagger\,.
\end{align}
The unitary mixing matrix $V_{\mathrm{CKM}}$ is characterized by 3 angles and 1 phase. The angles
describe flavor-changing transitions and the phase is responsible for the violation of \CP.
The standard parametrization of $V_{\mathrm{CKM}}$ is
\begin{align}
V_{\mathrm{CKM}} = \left(\begin{array}{ccc}
		c_{12} c_{13} & s_{12} c_{13} & s_{13} e^{-\I \delta} \\
		-s_{12} c_{23}-c_{12} s_{13} s_{23} e^{\I \delta} & c_{12} c_{23}-s_{12} s_{13} s_{23} e^{\I \delta} & c_{13} s_{23} \\
		s_{12} s_{23}-c_{12} s_{13} c_{23} e^{\I \delta} & -c_{12} s_{23}-s_{12} s_{13} c_{23} e^{\I \delta} & c_{13} c_{23}
	\end{array}\right)\,,
	\label{PMNS}
\end{align}
where $s_{ij}\equiv\sin\theta_{ij}$, $c_{ij}\equiv\cos\theta_{ij}$ ($0\le\theta_{ij}\le\pi/2$) and $0\le\delta< 2\pi$.

A last effect of the basis change concerns part of ${\cal L}_{gauge}$, the so-called topological QCD term
\begin{align}
\frac{g_S^2\theta}{32\pi^2}F^{a3}_{\mu\nu}\tilde F^{a3\mu\nu}\,,
\end{align}
where $F^{a3}_{\mu\nu}$ $(a=1,...,8)$ is the field strength of the SU(3) group and $\tilde F^{a3\mu\nu}=\frac12~\epsilon^{\mu\nu\rho\sigma}F^{a3}_{\rho\sigma}$ is the dual field strength tensor. This term is a total divergence but, on account of the existence of Yang-Mills instanton field configurations, can affect physical quantities. After the change of basis in the fermion sector, the parameter $\theta$ is replaced by
\begin{align}
\bar\theta=\theta+\arg\det Y^u+\arg\det Y^d\,.
\end{align}
It can be shown that the combination $\bar\theta$ is a physical quantity, invariant under field redefinitions,
providing an additional source of \CP violation beyond the $\delta$ phase in the quark mixing matrix $V_{\mathrm{CKM}}$.
 
In summary, the flavor sector of the SM is described by fourteen free parameters, originating from the matrices of Yukawa couplings $Y^{u,d,e}$ - in a one-to-one correspondence with the physical quantities $m^{u,d,e}_i$ ($i=1,2,3$),  $\theta_{12}$,   $\theta_{13}$,  $\theta_{23}$, $\delta$ and $\bar\theta$. All masses are proportional to the VEV $v$ of the Higgs field. In the limit $Y^{u,d,e}=0$, fermions are massless and the classical Lagrangian of the theory enjoy¿s a global symmetry under the group $\U3^5$, acting as an independent unitary transformation in generation space within each multiplet $f_i=(q_{Li},u_{Ri},d_{Ri},l_{Li},e_{Ri})$, and providing an example of chiral symmetry.
\vskip 0.5 cm
\paragraph{The GIM mechanism}
\begin{figure}[h!]
\caption{Diagrams for the transition $\bar K^0\to \mu^+\mu^-$. $\theta_C$ is the Cabibbo angle, equal to $\theta_{12}$
in the limit $\theta_{13}=\theta_{23}=0$.}
\vskip 0.2 cm
\label{GIM}
\centering
\includegraphics[width=.85\textwidth]{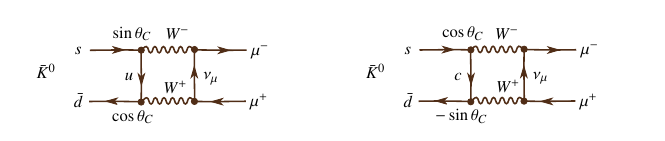}
\end{figure}
Only charged current interactions, mediated by the gauge bosons $W^\pm$, can produce transitions where the flavor of a particle changes, thanks to the quark mixing matrix $V_{CKM}$. At the classical level, neutral currents, mediated by the gauge bosons $Z$, involve particles that do not change neither their electric charge nor their flavor during the process.
FCNC can arise at the quantum level from the virtual exchange of the gauge bosons $W^\pm$, but the induced processes are extremely rare and, within the SM, are suppressed by the GIM mechanism~\cite{Glashow:1970gm}. An example of how the GIM mechanism works is shown in Fig.~\ref{GIM}.
The amplitude for the transition
$\bar K^0\to \mu^+\mu^-$ is proportional to $G_F (G_F M^2)$, where $G_F=1/(\sqrt{2}v^2)$ is the Fermi constant and $M$ is some mass parameter, combination of the masses of the particles circulating in the loop. The diagrams contributing to the amplitude are shown in Fig.~\ref{GIM}, assuming two fermion generations. Due to the unitarity of the mixing matrix,
the quark $\bar u$ and the quark $\bar c$ have mutual orthogonal couplings to the pair $(s,d)$.
For equal masses of $u$ and $c$ quarks the total amplitude vanishes. For different $u$ and $c$ masses, the combination $G_F M^2$ effectively becomes $G_F(m_c^2-m_u^2)$, providing a strong suppression factor. Any fundamental theory aiming to explain the origin of
fermion masses and mixing angles should avoid new sources of FCNC that might spoil this suppression.  
\subsection{Neutrino masses}\label{sec2:subsec2}
\begin{figure}[h!]
	\centering
	\includegraphics[width=12cm,height=3.5cm]{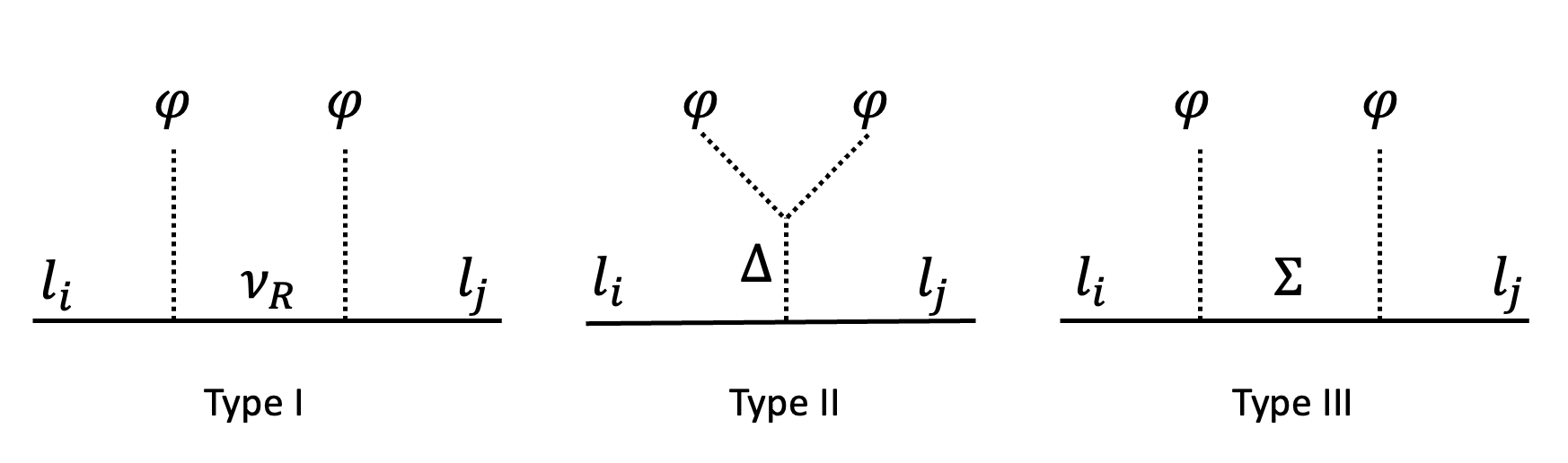}
	\caption{Feynman diagrams for the three types of seesaw.}
	\label{fig:see-saw}
\end{figure}

The SM does not account for neutrino masses and lepton mixing angles, well established from neutrino oscillation experiments. An extension of the SM is required to achieve consistency with the data. 
\paragraph{Majorana neutrino masses}
A minimal description of neutrino masses is obtained by relaxing the renormalizability requirement. Up to flavor combinations, there is a single dimension-five gauge-invariant operator within the field content of the SM, the Weinberg operator:
\begin{align}
\label{weinberg}
\delta{\cal L}_\nu=-\frac{1}{\Lambda}(\varphi l_{Li})^T \mathcal C ~W^\nu_{ij}~(\varphi l_{Lj}) +h.c.
\,,
\end{align}
where $\mathcal C$ is the charge conjugation matrix, and $\Lambda$ is a mass parameter. This operator delivers neutrino mass terms and interaction terms involving neutrinos and the Higgs boson. For generic values of the parameters $W^\nu_{ij}$, it violates both the individual lepton numbers $\mathrm{L}_e$, $\mathrm{L}_\mu$ and $\mathrm{L}_\tau$, and the total lepton number $\mathrm{L}=\mathrm{L}_e+\mathrm{L}_\mu+\mathrm{L}_\tau$. It adds 9 physical quantities
to the theory. The complex symmetric matrix $W^\nu$ is described by 6 absolute values and 6 phases. Performing the same change of basis done in the SM case, now the transformations of the group $\U1_{\mathrm{L}_e}\x\U1_{\mathrm{L}_\mu}\x\U1_{\mathrm{L}_\tau}$ do not leave the Lagrangian unchanged and can be
used to remove 3 phases. We end up with 6 absolute values and 3 phases as physical parameters.
In the new basis the matrix $W^\nu$ is diagonal and positive definite and, in the unitary gauge, $\delta{\cal L}_\nu$ becomes
\begin{align}
\delta{\cal L}_\nu=-\frac12\left(1+\frac{h}{v}\right)^2 \nu_{Li}^T C m^\nu_i\nu_{Li}+h.c.\,,
\end{align}
where $m^\nu_i=\frac{v^2}{\Lambda}W^\nu_{ii}$ are the 3 neutrino masses, called Majorana masses. Moreover, the charged currents $J^\pm_\mu$ read
\begin{align}
\label{Jplus}
J^{+\mu}=&~\bar d_L \gamma^\mu V_{\mathrm{CKM}}^\dagger u_L+ \bar e_L  \gamma^\mu U_{\mathrm{PMNS}} \nu_L\,,\qquad
J^{-\mu}=(J^{+\mu})^\dagger\,.
\end{align}
The unitary mixing matrix $U_{\mathrm{PMNS}}$ is characterized by 3 angles and 3 phases. The angles
describe flavor-changing transitions and the phases are responsible for the violation of \CP.
The standard parametrization of the mixing matrix $U_{\mathrm{PMNS}}$ is
\begin{align}
	U_{\text {PMNS }}= \left(\begin{array}{ccc}
		c_{12} c_{13} & s_{12} c_{13} & s_{13} e^{-\I \delta} \\
		-s_{12} c_{23}-c_{12} s_{13} s_{23} e^{\I \delta} & c_{12} c_{23}-s_{12} s_{13} s_{23} e^{\I \delta} & c_{13} s_{23} \\
		s_{12} s_{23}-c_{12} s_{13} c_{23} e^{\I \delta} & -c_{12} s_{23}-s_{12} s_{13} c_{23} e^{\I \delta} & c_{13} c_{23}
	\end{array}\right) P\,,
	\label{PMNS}
\end{align}
where $s_{ij}\equiv\sin\theta_{ij}$, $c_{ij}\equiv\cos\theta_{ij}$ ($0\le\theta_{ij}\le\pi/2$), and the matrix $P$ contains the Majorana phases $\eta_{1,2}$:
\begin{align}
   P=\left(
   \begin{array}{ccc}
    	e^{\I \eta_1 }&0&0\\
	0& e^{\I \eta_2} &0\\
	0&0&1
	\end{array}\right)\,.
\end{align}
The phase $\delta$ can vary in the interval from $0$ to $2\pi$, while $\eta_{1,2}$ vary from $0$ to $\pi$. Of course, $\theta_{ij}$ and $\delta$ are distinct from the analogous parameters of the quark mixing matrix. Majorana neutrino masses are qualitatively different from charged fermion (Dirac) masses. Beyond violating L by two units, they are proportional to the combination $v^2/\Lambda$, which depends on an additional mass scale.
\paragraph{The seesaw mechanism}
Through the seesaw mechanism, we can reproduce the Weinberg operator starting from a theory
containing additional heavy particles~\cite{Minkowski:1977sc,Yanagida:1979as,GellMann:1980vs,Mohapatra:1979ia,Glashow:1979pj,Schechter:1980gr,Magg:1980ut}. The Weinberg operator can be generated via a tree-level exchange of heavy particles in three different ways, as illustrated in Fig.~\ref{fig:see-saw}.
\paragraph{Type-I seesaw}
In type-I seesaw, a set of fermions $\nu_R$, invariant under the SM gauge interactions
is added. In this extended theory, there is a new term in the Yukawa Lagrangian: $-\bar \nu_{Ri} Y^\nu_{ij}~ \varphi~ l_{Lj}+h.c.$ Moreover, a renormalizable mass term for right-handed neutrinos $\nu_R$ is allowed by gauge-invariance: $-\frac12 \bar\nu_{Ri}C M_{Rij}\bar\nu^T_{Rj}+h.c.$, with a typical mass scale not necessarily related to the electroweak breaking scale $v$. 
Assuming such a scale to be much bigger than $v$, the new particles cannot be produced at the presently accessible energies, and the fields $\nu_{Ri}$ only occur as virtual states in physical amplitudes. The tree-level diagrams where 
$\nu_{Ri}$ are exchanged lead to the Weinberg operator at low energies, with the identification $W^\nu=-Y^{\nu T} \Lambda M_R^{-1} Y^\nu$. 
\paragraph{Type-III seesaw}
In type-III seesaw, the role of $\nu_R$ is played by an \SU2 triplet fermions $\Sigma$
with vanishing hypercharge. The Weinberg operator obtained via the tree-level exchange of $\Sigma$ is formally identical to the one of the type-I seesaw, where $M_R$ now denotes the $\Sigma$ mass matrix.
\paragraph{Type-II seesaw}
Finally, in type-II seesaw, the new particles consist of a scalar multiplet $\Delta$, 
\SU2 triplet with hypercharge $+1$. It has a renormalizable interaction with lepton doublets: $- W^{II}_{ij}( l_{Li})^T C ~\Delta ~(l_{Lj}) +h.c.$ The scalar potential contains interaction terms between $\Delta$ and $\varphi$, leading to a VEV for $\Delta$ of order $v^2/M_\Delta$, where $M_\Delta$ is the $\Delta$ mass. The corresponding Weinberg operator
has couplings proportional to $W^\nu=-W^{II} \Lambda M_\Delta^{-1}$.

\paragraph{Dirac neutrino masses}
The violation of the total lepton number L has not yet been established, and we can contemplate a scenario where
L is conserved, at least at the classical level. Neutrino masses, named Dirac as opposed to Majorana, can be described by adding to the SM field content 3 right-handed neutrinos $\nu_{Ri}$, and imposing by hand the conservation of L. In this framework, neutrino masses 
arise by the new term: $-\bar \nu_{Ri} Y^\nu_{ij}~ \varphi~ l_{Lj}+h.c.$ in the Yukawa Lagrangian. In the basis where
$Y^\nu$ is diagonal and positive definite, we have $m^\nu_i=\frac{v}{\sqrt{2}}Y^\nu_{ii}$. The charged current
is modified as in Eq.~(\ref{Jplus}), with a mixing matrix $U_{\mathrm{PMNS}}$ that now depends on 3 angles and 1 phase,
as for the quark sector. 
\subsection{The data}\label{sec2:subsec3}
\begin{figure}[h!]
	\centering
	\includegraphics[width=12cm,height=5cm]{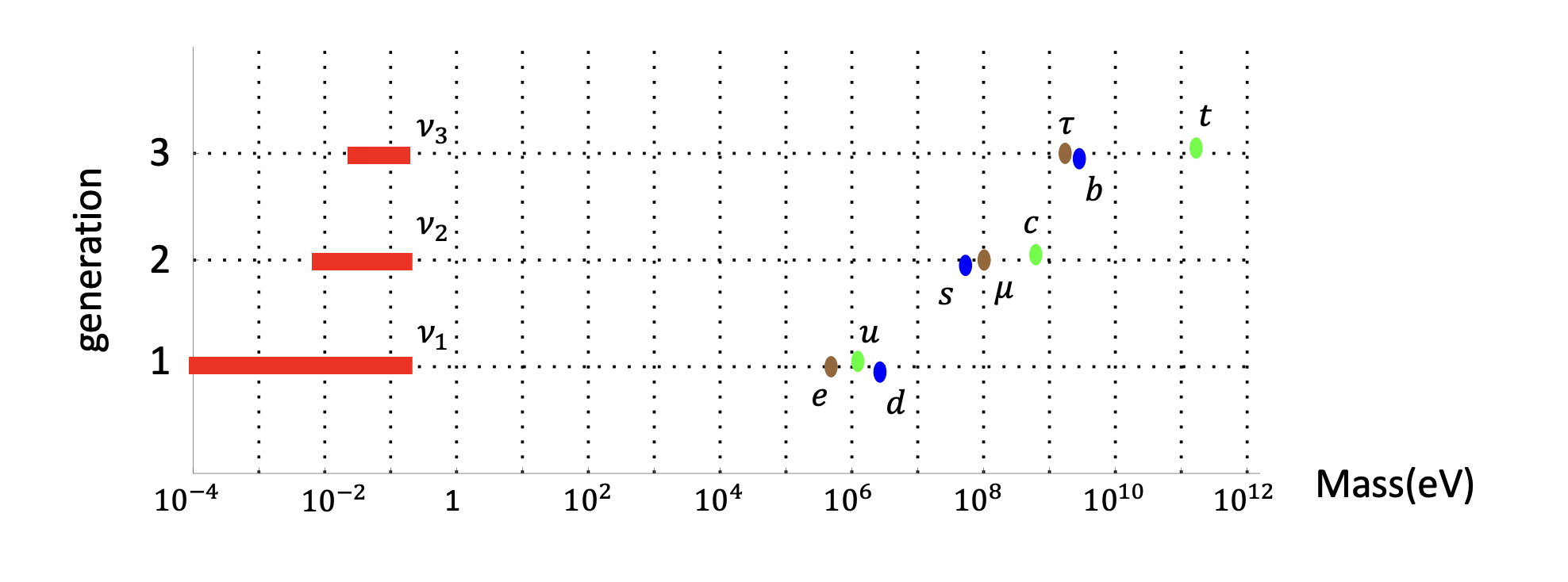}
	\caption{Hierarchical structure of fermion masses. Only upper bounds for neutrino masses exist. The figure assumes a normal ordering for neutrino masses. Masses are evaluated at the scale $m_Z$, see Table 4.}
	\label{fig1}
\end{figure}

In Fig.~\ref{fig1}, fermion masses evaluated in eV units are displayed on a logarithmic scale. Neutrino masses, not individually known,
appear to be special, their typical scale being separated by several orders of magnitude from the charged fermion masses.
It is also manifest the hierarchical structure of the charged fermion masses. Each generation of charged fermions has masses clustered in a relatively small interval, centered around few MeV for the first generation, 
between 0.1 GeV and 1 GeV for the second generation, and between 1 GeV and about 170 GeV for the third generation.
The intergenerational hierarchy is similar for charged leptons and down-type quarks, and is more pronounced for up-type quarks. At first sight, there is no evidence for an analogous hierarchy among neutrino masses, which appear anarchical compared to the ordered charged fermion spectrum.  
Also quark and lepton mixing angles turn out to be very different. The quark mixing matrix is nearly diagonal, in first approximation. It is dominated by the mixing
among the first two generations, characterized by the Cabibbo angle of order 0.2 rad. The off-diagonal entries of the quark mixing matrix outside the first-two-generation block are much smaller. On the contrary, most of the entries of the lepton mixing matrix are of the same order, with the exception of $U_{e3}$, whose absolute value is smaller.
\begin{table}[h!]
\parbox{.45\linewidth}{
\TBL{\caption{\label{tab:chargedFermions}Charged fermion masses, quarks mixing angles and \CP phase $\delta^q$.}}{%
		\begin{tabular}{ll}
		\toprule
$m_u(2~\mathrm{GeV})$ & $2.16\pm 0.07$ MeV\\
$m_d(2~\mathrm{GeV})$ & $4.70\pm 0.07$ MeV\\ 
$m_s(2~\mathrm{GeV}) $ & $93.5\pm 0.8$ MeV\\
\colrule
$m_c(m_c)$ & $1.273\pm 0.004$ GeV\\
$m_b(m_b)$ & $4.183\pm 0.007$ GeV \\
$M_t $ & $172.57\pm0.29$ GeV\\
\colrule
$M_e$ & $0.51099895000\pm 0.00000000015$ MeV\\
$M_\mu$ & $105.6583755\pm 0.0000023$ MeV\\
$M_\tau $ & $1776.93\pm 0.09$ MeV\\
\colrule
$\sin\theta^q_{12}$ & $0.22501\pm 0.00068$\\[0.1 cm]
$\sin\theta^q_{23} $ & $0.04183^{+0.00079}_{-0.00069}$\\[0.1 cm]
$\sin\theta^q_{13} $ & $0.003732^{+0.000090}_{-0.000085}$,\\[0.1 cm]
\colrule
$\delta^q $ & $1.147\pm 0.026$\\
	\botrule
	\end{tabular}}{}
	\vspace{9mm}
}
\hspace{0.5 cm}
\parbox{.4\linewidth}{
\TBL{\caption{\label{tab:Neutrinos}Neutrino mass squared differences, lepton mixing angles and \CP phase $\delta^l$.}}{%
		\begin{tabular}{lcl}
		\toprule
$\delta m^2\equiv m_2^2-m_1^2$ & NO, IO &$(7.37^{+0.15}_{-0.16})\times 10^{-5}$ eV$^2$\\[0.1 cm]
$|\Delta m^2|\equiv \left| m_3^2-\frac12(m_1^2+m_2^2)\right|$ & NO & $(2.495\pm 0.020)\times 10^{-3}$ eV$^2$\\ [0.1 cm]
& IO & $(2.465^{+0.020}_{-0.021})\times 10^{-3}$ eV$^2$\\ [0.1 cm]
\colrule
$\sin^2\theta^l_{12}$ & NO, IO & $0.303^{+0.014}_{-0.012}$\\[0.1 cm]
$\sin^2\theta^l_{23} $ & NO & $0.473^{+0.023}_{-0.013}$\\[0.1 cm]
& IO & $0.545^{+0.015}_{-0.017}$\\[0.1 cm]
$\sin^2\theta^l_{13} $ & NO &$0.0223^{+0.0004}_{-0.0006}$\\[0.1 cm]
& IO &$0.0223^{+0.0007}_{-0.0004}$\\[0.1 cm]
\colrule
$\delta^l $ & NO & $3.77^{+0.53}_{-0.41}$\\[0.1 cm]
& IO & $4.65^{+0.41}_{-0.38}$\\[0.1 cm]
\colrule
$m_\beta\equiv\sqrt{\sum_i |U_{ei}|^2~m_i^2}$ && $<0.5~\mathrm{eV}~~~~~~~~(2\sigma)$\\ [0.1 cm]
$\Sigma\equiv m_1+m_2+m_3$ && $<(0.07\div 0.60)~\mathrm{eV}$\\ [0.1 cm]
$m_{\beta\beta}\equiv|\sum_i U_{ei}^2~m_i|$&& $< 0.086~\mathrm{eV}~~~~~~~~(2\sigma)$\\ [0.1 cm]
	\botrule
	\end{tabular}}{}
}	
\end{table}

Experimental values for fermion masses are given  in terms of either the pole mass $M$ or the running mass
$m(\mu)$. The pole mass $M$ is the real part of the pole of the particle propagator, in momentum space.
The pole mass is gauge invariant, independent from the renormalization scheme, and can be directly extracted from the kinematics 
of particles in the final state of a process. 
The running mass $m(\mu)$ is defined as the mass parameter renormalized according to a specific scheme, thus
making reference to a renormalization scale $\mu$.
In a given scheme, $M$ and $m(\mu)$ are related by a well-defined equation of the type:
\begin{align}
\label{pole}
M=m(\mu)\left(1+\sum_{n=1}^\infty c_n(\mu)\right)\,,
\end{align}
where $c_n(\mu)$ is evaluated at the loop-order $n$ and depends on the renormalized parameters of the theory.
Once the scale $\mu$ is given, and the above relation
is known, it is a matter of convenience to make use of the pole or the running mass.
However, for all quarks except the heaviest one, the series in the above expression does not converge
due to nonperturbative corrections of the order of $\Lambda_{QCD}/\mu$, with $\Lambda_{QCD}$ a typical QCD scale of order a few hundred MeV. Moreover, up, down, strange, charm, and bottom quarks all
hadronize, that is become part of a meson or baryon, on a timescale $1/\Lambda_{QCD}$ and their pole mass cannot be 
extracted from a direct measurement. The top quark instead decays before it has time to hadronize. For all quarks but the top one, masses refer to running masses, usually evaluated in the $\overline{MS}$ scheme.

The values of charged fermion masses and quark mixing angles from the latest PDG review \cite{ParticleDataGroup:2024cfk}, are shown in Table~\ref{tab:chargedFermions}. In Table~\ref{tab:Neutrinos}, the lepton mixing angles, the phase $\delta^l$ and combinations $(\delta m^2,\Delta m^2)$ of neutrino masses are shown. They have been derived from a global fit to neutrino oscillation experiments~\cite{Capozzi:2025wyn}, which are only sensitive to the mass squared differences $\delta m^2\equiv m_2^2-m_1^2>0$
and $\Delta m^2\equiv m_3^2-\frac12(m_1^2+m_2^2)$. Moreover, the sign of $\Delta m^2$ has not yet been determined, resulting in
a twofold ambiguity for the mass ordering: normal ordering (NO) if $\Delta m^2>0$, and inverted ordering (IO) if $\Delta m^2<0$.

Also shown in Table~\ref{tab:Neutrinos} are an upper limit on the combination $m_\beta$, derived from the analysis of the endpoint of the electron energy spectrum in tritium beta decay, and an (uncertain) upper limit on the sum $\Sigma$ of neutrino masses derived from cosmology.
An open question is whether the total lepton number L is violated or not. The most sensitive process to L violation is neutrinoless double beta decay, a nuclear transition of the type $(A,Z)\to (A,Z+2)~e^- e^-$, where L is violated by two units. If neutrinos have Dirac masses, L is conserved and neutrinoless double beta decay is forbidden. If neutrinos have Majorana masses, from the neutrinoless double beta decay 
we can have access to the combination $m_{\beta\beta}$ of Table~\ref{tab:Neutrinos}, where also Majorana phases enter. At the moment,
we have only lower bounds on the half-life of these transitions, resulting in the approximate bound on $m_{\beta\beta}$ shown in Table~\ref{tab:Neutrinos}.
\begin{table}[h!]
	\TBL{\caption{\label{tab:chargedFermionsAtScales}Quark and charged lepton masses renormalized at two different scales, from ref.~\cite{Huang:2020hdv}. Experimental inputs slightly differ from those of Tables~\ref{tab:chargedFermions} and~\ref{tab:Neutrinos}.}}{%
		\begin{tabular*}{\columnwidth}{@{\extracolsep\fill}ccccccc@{}}
		\toprule
$\mu$ & $m_u(\mu)$ & $m_d(\mu)$ & $m_s(\mu)$\\
\colrule
$m_Z$ & $(1.23\pm0.21)~ \mathrm{MeV}$ & $(2.67\pm0.19)~ \mathrm{MeV}$ & $(53.16\pm4.61)~ \mathrm{MeV}$\\[0.1cm]
$10^{12}~\mathrm{GeV}$ & $(0.56\pm0.10)~ \mathrm{MeV}$ & $(1.24\pm0.09)~ \mathrm{MeV}$ & $(24.76\pm2.17)~ \mathrm{MeV}$\\[0.1cm]
\colrule
$\mu$ & $m_c(\mu)$ & $m_b(\mu)$ & $m_t(\mu)$\\
\colrule
$m_Z$ & 
$(0.620\pm0.017)~ \mathrm{GeV}$ & $(2.839\pm0.026)~ \mathrm{GeV}$ & $(168.26\pm0.75)~ \mathrm{GeV}$ \\[0.1cm]
$10^{12}~\mathrm{GeV}$ & 
$(0.283 \pm0.009)~ \mathrm{GeV}$ & $(1.194\pm0.015)~ \mathrm{GeV}$ & $(85.07\pm0.89)~ \mathrm{GeV}$\\
\colrule
$\mu$ & $m_e(\mu)$ & $m_\mu(\mu)$ & $m_\tau(\mu)$ \\
\colrule
$m_Z$ & $(0.48307\pm0.00045)~ \mathrm{MeV}$ & $(0.101766\pm0.000023)~ \mathrm{GeV}$ & $(1.72856\pm0.00028)~ \mathrm{GeV}$   \\[0.1cm]
$10^{12}~\mathrm{GeV}$ & $(0.48388\pm0.00139)~ \mathrm{MeV}$ & $(0.101936\pm0.000277)~ \mathrm{GeV}$ & $(1.73194\pm0.00466)~ \mathrm{GeV}$   \\[0.1cm]
	\botrule
	\end{tabular*}}{}
\end{table}

If the SM is interpreted as the low-energy approximation of some fundamental theory whose characteristic scale is $\Lambda_F\gg v$,
the relevant values of fermion masses and mixing angles in the new theory are the renormalized ones, evaluated at the scale $\mu=\Lambda_F$.
The scale dependence and the evolution of the parameters from low-energies up to the scale $\Lambda_F$ are governed by a renormalization group equation. For fermion masses $m(\mu)$, this equation schematically reads 
\begin{align}
\mu\frac{d}{d\mu} m(\mu) =\gamma_m~ m(\mu)\,,
\end{align}
where $\gamma_m$, the anomalous dimension of $m$, depends on the renormalized parameters of the theory. This equation can be derived from Eq.~(\ref{pole}), by differentiating both sides with respect to $\log\mu$ and recalling that the pole mass $M$ does not depend on $\mu$. 

Predictions about the parameters of the flavor sector, derived from a theory with typical scale $\Lambda_F$, should 
be compared with the experimental values extrapolated at $\mu=\Lambda_F$, which can considerably vary from those listed in Tables~\ref{tab:chargedFermions} and~\ref{tab:Neutrinos}.
In Table~\ref{tab:chargedFermionsAtScales} we show charged fermion masses evaluated at two illustrative scales. To realize the relevance of
such scale dependence, consider the Koide empirical formula~\cite{Koide:1983qe},
\begin{align}
\frac{\left(\sqrt{m_e}+\sqrt{m_\mu}+\sqrt{m_\tau}\right)}{\sqrt{m_e+m_\mu+m_\tau}}=\sqrt{\dd\frac{3}{2}}~~~~,
\label{koide}
\end{align}
which relates the three charged fermion masses. If we use as inputs $m_e=M_e$ and $m_\mu=M_\mu$, adopting the central values of Table~\ref{tab:chargedFermions}, from this relation we get $m_\tau=1776.97$ MeV, in excellent agreement with the value $M_\tau$. However, if the three masses are believed to
originate from a theory at a common scale $\Lambda_F$, it is more adequate to check Koide's formula using running masses evaluated at $\Lambda_F$. For example, if $\Lambda_F=10^{12}$ GeV, taking as inputs $m_e=m_e(\Lambda_F)$ and $m_\mu=m_\mu(\Lambda_F)$
from Table~\ref{tab:chargedFermionsAtScales}, we derive $m_\tau=1711.58$ MeV, smaller than $m_\tau(\Lambda_F)$ by about 20 MeV, a discrepancy much bigger than the experimental uncertainty. This example relies on an extrapolation of the SM at energies way bigger than those currently explored.
New Physics could affect the running of the parameters and its effects should be taken into account to compare
a prediction like the one in Eq.~(\ref{koide}) with the data. 
\section{In search of an organizing principle}\label{search}
With upcoming experiments, this broad picture could be expanded by unforeseen discoveries, possibly arising from different areas of research. The issue of family structures and fermionic masses is deeply connected to the origin of the electroweak scale and its stability in relation to the Planck scale. Valuable perspectives on this matter could emerge from both current particle accelerators and those still in development. Many theoretical frameworks aiming to explain the existence of families suggest, to some degree, an enhancement of FCNC beyond the expectations of the GIM mechanism. Numerous experiments designed to probe extremely rare processes within the SM are either in progress or being planned. Additionally, the vast majority of matter in the universe takes the form of dark matter, whose fundamental nature remains a mystery. If dark matter is associated with an undiscovered particle, there may be a direct link to the origin of families and their interactions. The search for dark matter stands at the forefront of advancements in both particle physics and astrophysics today.

Although significant future breakthroughs remain possible, precise measurements of fermionic masses and mixing angles have so far provided a consistent overall picture, sufficient to raise profound theoretical questions. Why do exactly three families exist? What determines the hierarchy of charged fermion masses, where the third generation is significantly heavier than the second, which in turn is much heavier than the first? Why are quark mixing angles so small, while leptonic mixing angles do not follow the same pattern observed in the quark sector?

Within the SM, each mass, mixing angle, and observable phase represents a free parameter that, by definition, cannot be derived from the model itself. The theory permits any values for these parameters in principle. It is crucial to acknowledge that reducing these parameters to a more fundamental or minimal set may not be feasible and could be a misguided pursuit. The history of physics provides relevant lessons in this regard. A well-known example is the arrangement of planetary orbits in the solar system, which was once considered a fundamental mystery. Kepler himself proposed a solution to this issue in his work Mysterium Cosmographicum. However, modern understanding suggests that these orbital relationships are not an intrinsic property of Nature, as our solar system holds no special status in the universe. A similar perspective could apply to the parameters governing fermion masses in the SM. Nonetheless, if one assumes that these parameters should, in principle, be calculable, then it is necessary to go beyond the SM~\cite{Weinberg:1977hb,Peccei:1997mz}. Below, we will examine various approaches that seek to determine the masses of quarks and leptons using a reduced set of independent parameters.
\section{Grand Unified Theories}\label{sec4}
Among the many unsolved questions in particle physics, one of the most intriguing is the origin of the SM gauge group and the quantum numbers of elementary particles. A convincing, yet still unconfirmed, explanation comes from embedding
the SM gauge group into a larger simple group~\cite{Georgi:1974sy}.
Just as electromagnetism can be understood as the low-energy remnant of a gauge theory based on \SU2\x\U1, the electroweak and strong interactions may likewise arise as the low-energy manifestation of a more fundamental unified theory. This underlying symmetry would become fully apparent only at extremely high energies, well beyond the current experimental reach.
The SM gauge group $G_{SM}$ can be embedded within larger, simple groups such as \SU5, \SO{10}, or \E6. A key consequence of this embedding is the unification of the three gauge couplings $g_S$, $g$, and $\sqrt{5/3}g'$ at a particular energy scale, $M_{GUT}$. The factor $\sqrt{5/3}$ appears due to the proper normalization of the \U1 hypercharge generator, ensuring consistency with the other SM generators.
Because gauge couplings evolve with energy, the values of $g_S$, $g$, and $\sqrt{5/3}g'$ tend to converge as the energy scale increases. In fact, in the SUSY extension of the SM, these running couplings become equal at $M_{GUT} \approx 10^{16}$ GeV, provided that the superpartner masses lie in the range $\mathcal{O}(1 - 10)$ TeV. In this framework, the different values of the gauge couplings observed at low energies arise from the breaking of the grand unified symmetry at a very high energy scale, not far from the Planck scale, $M_{Pl} \approx 2.4 \times 10^{18}$ GeV. Additionally, the quantum numbers of quarks and leptons - left unexplained within the SM - emerge naturally from the simplest GUT representations.
\subsection{SU(5) GUTs}
Several SM multiplets merge into a few irreducible representations of GUTs. A single SM family is embedded in two representations of \SU5:
$\rep{10} = (q_L, u_R^c, e_R^c)$ and $\crep{5} = (l_L, d_R^c)$, where the superscript $c$ stands for charge conjugation: $\psi_R^c=C (\bar \psi_R)^T$. In this framework, the distinction between quarks and leptons becomes meaningless, hinting at underlying relationships among their parameters.
For example, in \SU5, the Higgs electroweak doublet $\varphi$ can be minimally incorporated into the representation $\rep5_H$, which also contains an additional spin-zero color triplet. The Yukawa interactions in this minimal GUT scenario are given by:
\begin{align}
\label{minsu5}
-\rep{10}_i~Y^u_{ij}~ \rep{10}_j~ \rep5_H - \rep{10}_i~ Y^d_{ij}~ \crep{5}_j~ \crep{5}_H\,,
\end{align}
leading to charged fermion mass matrices in a generic basis: 
$m_u = Y^u \langle \rep5_H \rangle$ and $m_e = m_d^T = Y^d \langle \crep{5}_H \rangle$. This structure implies that, at the GUT scale, charged lepton and down-type quark masses are equal. However, a significant extrapolation from the GUT scale down to the electroweak scale is necessary to compare this prediction with the data.
In the SUSY version of the SM, which supports gauge coupling unification, this extrapolation involves renormalization group (RG) evolution and threshold corrections due to the masses of SUSY partners. SUSY also necessitates two Higgs multiplets, $\rep5_H$ and $\crep{5}_H$, where the ratio of their VEVs, $\tan\beta = v^u_5 / v^d_5 = \langle \rep5_H \rangle / \langle \crep{5}_H \rangle$, becomes a crucial parameter in the analysis. The relation $m_\tau (M_{GUT}) = m_b (M_{GUT})$ holds in a certain region of parameter space, strongly dependent on $\tan\beta$. However, the relation $m_s (M_{GUT}) / m_d (M_{GUT}) = m_\mu (M_{GUT}) / m_e (M_{GUT})$ is approximately scale-invariant and disagrees with experimental data. From Table~\ref{tab:chargedFermionsAtScales}, we obtain $m_s (m_Z) / m_d (m_Z) \approx 20$, to be compared with $m_\mu (m_Z) / m_e (m_Z) \approx 210$. Despite this discrepancy, it is remarkable that such a minimal theory qualitatively captures the similarity between charged lepton and down-type quark mass ratios.

This minimal framework can be modified in several ways to align GUT-scale predictions with experimental data. One approach is to introduce a Higgs multiplet in the $\rep{45}$ representation of \SU5, modifying the Yukawa interactions~\cite{Georgi:1979df}.
In particular, the combination
\begin{align}
- \rep{10}_i~ (Y^{5}_{ij}~\crep5_H+Y^{45}_{ij}~\rep{45}_H)~ \crep5_j\,
\end{align}
leads to $m_e=v^d_5Y^{5}-3v^d_{45}Y^{45}$ and $m_d^T=v^d_5Y^{5}+v^d_{45}Y^{45}$,  providing sufficient flexibility to modify the incorrect mass relations. Here, $v^d_{45}$ is the VEV of the Higgs doublet with hypercharge $-1/2$ in $45_H$. The data can be well approximated with:
\begin{align}
\label{GJ}
v^d_5 Y^{5}=
\left(
\begin{array}{ccc}
0 & A & 0\\
A & 0 & 0\\
0 & 0 & C
\end{array}
\right), ~~~~~~
v^d_{45} Y^{45}=
\left(
\begin{array}{ccc}
0 & 0 & 0\\
0 & B & 0\\
0 & 0 & 0
\end{array}
\right),
\end{align}
where $A \ll B \ll C$. This leads to the relations $m_\tau (M_{GUT}) = m_b (M_{GUT})$, $m_\mu (M_{GUT}) \approx 3 m_s (M_{GUT})$, and $m_d (M_{GUT}) \approx 3 m_e (M_{GUT})$. Furthermore, if the Cabibbo angle $\theta^q_{12}$ is dominated by the down-quark sector, then $\theta^q_{12} \approx \sqrt{m_d / m_s}$, a well-known relation proposed by Gatto, Sartori, and Tonin in the late 1960s~\cite{Gatto:1968ss}.

Another possibility is to assume that the minimal Yukawa interactions in Eq.~\eqref{minsu5} mainly determine third-generation masses, enforcing the prediction $m_\tau (M_{GUT}) \approx m_b (M_{GUT})$, while additional small contributions to the first and second generations arise from higher-dimensional operators~\cite{Ellis:1979fg}. Since $M_{GUT}$ is only a few orders of magnitude below the Planck scale $M_{Pl}$, gravitational effects could play a non-negligible role in GUTs. At the GUT scale, these effects can be described by non-renormalizable operators suppressed by powers of $M_{Pl}$. Even with a minimal field content, such operators can be constructed using the adjoint Higgs representation $24_H$, required to break SU(5) down to the SM gauge group $G_{SM}$ and developing
a VEV of order $M_{GUT}$. An example is:
\begin{align}
 -\frac{1}{M_{Pl}} \rep{10}_i~ Y^{d'}_{ij}~ \crep{5}_j~ \rep{24}_H \crep{5}_H\,.
\end{align}
When SU(5) indices are contracted appropriately, this operator generates mass terms of order $v M_{GUT} / M_{Pl} \ll v$ for charged leptons and down-type quarks, modifying the relation $m_s (M_{GUT}) / m_d (M_{GUT}) = m_\mu (M_{GUT}) / m_e (M_{GUT})$.

Neutrino masses and lepton mixing angles can be addressed as in Section \ref{sec2:subsec2}. The generalization of the Weinberg operator is given by:
\begin{align}
\delta {\cal L}_\nu=-\frac{1}{\Lambda} (\rep5_H \crep{5}_i)^T C W^\nu_{ij} (\rep5_H \crep{5}_j)+h.c.
\end{align}
This operator can be realized through the seesaw mechanism if right-handed neutrinos $\bar{\nu}_R$, assigned to singlet representations of SU(5), are included. Their Yukawa couplings are given by $-\rep 1_i Y^\nu_{ij} \crep{5}_j \rep5_H + h.c.$ Right-handed neutrino masses $M_R$ are expected to be of order $M_{GUT}$ or higher. Assuming $M_R \approx M_{GUT}$ and Yukawa couplings $Y^\nu$ of order one, neutrino masses are estimated as $v^2 / M_{GUT} \approx 0.001$ eV, not very far from the square root of $|\Delta m^2|$ from Table~\ref{tab:Neutrinos}.

A puzzling question is how to simultaneously generate small quark mixing angles and large lepton mixing angles, given that quarks and leptons reside in the same irreducible representations.
In \SU5 this is possible thanks to the fact that quark electroweak doublets and lepton electroweak doublets
are part of different \SU5 representations.
The mixing matrices originate from the unitary transformations needed to move from the generic basis, where interactions are defined, to the mass basis, where mass matrices are diagonal and positive definite.  
The $V_{\mathrm{CKM}}$ matrix is sourced by the transformations done on $u_L$ and $d_L$ contained in $\rep{10}_i$, while
the matrix $U_{\mathrm{PMNS}}$ is obtained by transforming $\nu_L$ and $e_L$ in $\crep5_i$.
Thus the different features of $V_{\mathrm{CKM}}$ and $U_{\mathrm{PMNS}}$ can be explained if the required transformations involve, approximately, small(large) angles for the members of the $\rep 10_i$($\crep 5_i$) multiplets.
This picture implies potentially large mixing angles for $\bar d_R$ - the partner of $l_L$ in $\bar 5$ - which however are not observable at our energies
due to the peculiar structure of charged electroweak currents that only involve \SU2 doublets.
\subsection{SU(10) GUTs}
A distinctive feature of \SO{10}-based GUTs compared to \SU5 theories is the restoration of parity. The theory is expected to be left-right symmetric, at least at the GUT scale. Moreover, the 16-dimensional representation of \SO{10} accommodates an entire SM family along with a right-handed neutrino:
\begin{equation}
\rep{16} = (q_L, l_L, u_R^c, d_R^c, \nu_R^c, e_R^c)\,,
\end{equation}
a structure that is too remarkable to be coincidental.  
As in SU(5), minimal Yukawa interactions, which in principle offer the strongest predictive power, are insufficient to fully reproduce experimental data. The tensor product decomposition  
$\crep{16} \times \crep{16} = \rep{10} \oplus \rep{120} \oplus \crep{126}$
suggests a minimal embedding of the Higgs fields in the 10-dimensional representation, which contains two electroweak doublets. However, the corresponding Yukawa interactions,  
$\rep{16}_i Y^{10}_{ij} \rep{16}_j \rep{10}_H$,
are too restrictive and must be supplemented with additional terms. The most general renormalizable Yukawa interactions take the form
\begin{align}
- \rep{16}_i \left( Y^{10}_{ij} \rep{10}_H + Y^{120}_{ij} \rep{120}_H + Y^{126}_{ij} \crep{126}_H \right) \rep{16}_j + h.c.
\end{align}
In addition to SU(2) doublets, the $\crep{126}$ multiplet contains both a singlet and a triplet, which can develop VEVs $v_R$ and $v_L$, respectively.
In the simplest scenario, where $Y^{120}_{ij} = 0$, integrating out the right-handed neutrinos leads to the following mass relations:
\begin{align}
m_d &= v^d_{10} Y^{10} + v^d_{126} Y^{126},  & m_u &= v^u_{10} Y^{10} + v^u_{126} Y^{126}, \\
m_e &= v^d_{10} Y^{10} - 3 v^d_{126} Y^{126}, &
m_\nu &= v_L Y^{126} - \frac{v^2}{v_R} Y^{\nu T} (Y^{126})^{-1} Y^\nu, 
\end{align}
where $v Y^\nu \equiv  v^u_{10} Y^{10} -3 v^u_{126} Y^{126}$.
Here, $v^{u,d}_{10}$ and $v^{u,d}_{126}$ denote the VEVs of the doublets within $\rep{10}_H$ and $\crep{126}_H$, respectively.
The mass matrices of charged leptons and down-type quarks resemble those in \SU5 with the $\rep{45}_H$ representation, allowing to relax the rigid relations of the minimal model. The neutrino mass matrix receives two contributions: the first, proportional to $Y^{126}$, arises from the type-II seesaw mechanism, while the second stems from integrating out the right-handed neutrinos via the type-I seesaw mechanism, as discussed in Section \ref{sec2:subsec2}. Both mechanisms naturally generate small neutrino masses, as $v_L$ is typically of order $v^2 / v_R$, and $v_R$ can be much larger than the electroweak scale $v$.

Large mixing angles in the lepton sector can coexist with small quark mixing angles if neutrino masses are dominated by the first contribution~\cite{Bajc:2002iw}, leading to  
$m_\nu \propto m_d - m_e$.
If both $m_d$ and $m_e$ exhibit the following structure:
\begin{align}
\left(
\begin{array}{ccc}
\cdot & \cdot & \cdot \\
\cdot & B' & B \\
\cdot & B & C
\end{array}
\right),
\end{align}
with $B \approx B' \ll C$ and the dots representing smaller entries, the near equality between $m_b(m_{GUT})$ and $m_\tau(m_{GUT})$ results in the cancellation of the $C$ term in $m_d - m_e$, thereby allowing for a large lepton mixing angle in the 23 sector.  
\paragraph{Summary}  
GUTs may offer valuable insights into the underlying structure of fermion masses. Quarks and leptons are unified within common multiplets of the GUT group, leading to expected relations among the Yukawa couplings. Moreover, \SO{10} GUTs naturally incorporate the key ingredients of the seesaw mechanism --- namely, the existence of right-handed neutrinos and their large mass scales --- thus providing a compelling explanation for the smallness of neutrino masses.
Despite these appealing features, minimal GUT models are typically too constrained to accurately reproduce the observed fermion mass spectrum. While the masses of charged fermions can be accommodated by introducing additional parameters, this comes at the cost of reducing the model's predictive power. The hierarchical structure among different fermion generations remains unexplained and must be enforced by fine-tuning. Some predictive power can be retained by setting specific Yukawa couplings to zero, as in Eq.~\eqref{GJ}, but at this stage, such conditions appear arbitrary rather than emerging naturally from the theory.

\section{Fermion masses from quantum corrections}\label{sec5}
Beyond representing an indispensable tool to relate model predictions and experimental data, radiative corrections might
help explaining the hierarchical structure of fermion masses.
\subsection{Radiative fermion masses}\label{sec5dot1}
Since the fermions of the third generation are considerably heavier than those of the first two, it is reasonable to consider that the lighter masses might arise through radiative corrections from the heavier ones - an idea analyzed by Steven Weinberg in one of his last works~\cite{Weinberg:2020zba}. 
The core idea is that only the heaviest fermions, for example those of the third generation, obtain their masses at tree level. The fermions of the second generation, which are slightly lighter, acquire their masses through one-loop radiative corrections, leading to a suppression by a typical loop factor of approximately $1/(16\pi^2)\approx 10^{-2}$,
compared to the heaviest fermions. Meanwhile, the lightest fermions, like the up and down quarks and the electron, obtain their masses via two-loop radiative corrections, introducing an additional suppression factor of roughly $1/(16\pi^2)^2\approx 10^{-4}$
relative to the heaviest ones. This mechanism can naturally explain the observed hierarchy in the fermion mass spectrum without requiring small Yukawa couplings from the outset.

However, if only the third-generation fermions have nonzero tree-level masses
and we remain within the SM, radiative effects alone cannot generate the masses of the remaining fermions. To illustrate this point, consider the
simple case where all mass matrices share the pattern:
\begin{align}
\left(
\begin{array}{ccc}
0 & 0& 0\\
0 & 0 & 0 \\
0 & 0 & C
\end{array}
\right),
\end{align}
where $C$ depends on the specific charge sector, $u$, $d$ and $e$. Such a theory enjoys a chiral symmetry of the type $\U2^5$, where each \U2 acts independently on the first two generations of the multiplets $f_i=(q_{Li},u_{Ri},d_{Ri},l_{Li},e_{Ri})$. This symmetry, obeyed by all interactions of the theory, is maintained by perturbative corrections.  The above pattern is stable and fermion masses for the first two generations remain massless to any loop order.

To overcome this constraint, one must look beyond the SM and consider a theory in which the first two generations of fermions are massless at the tree level due to accidental reasons.
Moreover, for the theory to be predictive, masses arising in perturbation theory for particles that are massless
at tree level should be finite, i.e. free from divergences. 
This requirement is met in theories where
vanishing tree-level masses arise entirely as a consequence of gauge symmetry, particle content and renormalizability~\cite{Weinberg:1972ws,Georgi:1972hy}.
Indeed, consider a theory invariant under a gauge group $G$, possibly spontaneously broken down to a subgroup $H$. 
Suppose that some mass terms vanish for all possible renormalizable $G$-invariant Lagrangians, 
built from a given set of fields forming $G$-multiplets.
Then the higher-order corrections to this mass terms must be finite, because there are no counterterms available to absorb ultraviolet divergences in these corrections.

\begin{figure}[h!]
	\centering
	\includegraphics[width=.35\textwidth]{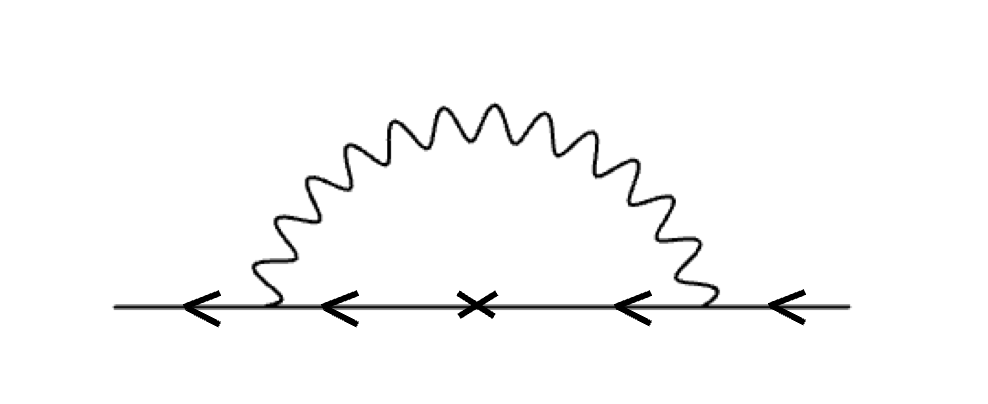}
	\caption{One-loop diagram contribution to fermion mass terms. Bosons and massive
fermions in the loop are understood to be in the mass eigenstate basis.}
	\label{diag}
\end{figure}

The group $G$ can be an extension of the SM gauge group,
such as $G=G_{SM}\times G_{NP}$, where $G_{NP}$ is some new group factor, and a convenient Higgs sector can
spontaneously break $G$  down to $\SU3\x\U1_{em}$.
For specific choices of $G$ representations of fermion and Higgs multiplets, 
it can happen that the gauge symmetry forces all renormalizable Yukawa interactions to enjoy the above $\U2^5$
symmetry, otherwise broken in other sectors of the theory. In this case $\U2^5$ is not a symmetry of the full theory,
but rather an accidental property of the Yukawa sector. Nothing prevents that the masses of the lighter fermions emerge from quantum corrections induced by the additional fields and the interactions which explicitly break the chiral symmetry, 
For example, a $\U2^3$ symmetry limited to the Yukawa quark sector arises if there are no scalars transforming
as $\bar u_{Ra} q_{Lb}$ and $\bar d_{Ra} q_{Lb}$ ($a,b=1,2$), or if these scalars exist but do not develop VEVs. Mass terms can then be generated by loop diagrams with bosons and massive fermions as the one in Fig.~\ref{diag}.

The third quark generation can become massive thanks to a mixing with heavy fermions~\cite{Balakrishna:1988ks}.
As an example, consider the up quark sector consisting of the $u$, $c$ and $t$ quarks and an additional color triplet $T$ with electric charge $-2/3$. 
Assume that particle content, gauge symmetry and renormalizability requires
that, after $G$ symmetry breaking, the most general mass term for $f_{R,L}=(u_{R,L},c_{R,L},t_{R,L},T_{R,L})$ reads $-\bar f_R {\cal M}^0_u f_L+h.c.$, where ${\cal M}^0_u$ is the following rank-2 matrix:
\begin{align}
{\cal M}^0_u=\left(
\begin{array}{cccc}
0&0&0&\mu'_1\\
0&0&0&\mu'_2\\
0&0&0&\mu'_3\\
\mu_1&\mu_2&\mu_3&M
\end{array}
\right)\,.
\end{align}
Consider the simple case of $G_{NP}=\,$U(1)$_X$.
The associated gauge interactions involving $f_{R,L}$
read, in a matrix notation
\begin{align}
- g_X X_\mu (\bar f_L X_L \gamma^\mu f_L+\bar f_R X_R \gamma^\mu f_R)\,,
\end{align}
where $X_\mu$ is the new gauge vector boson, $g_X$ the gauge coupling constant and $X_{L,R}$ are
diagonal matrices collecting the $X$ charges of $f_{R,L}$. The tree-level mass matrix is diagonalized by unitary transformations
$f_{L,R}\to {\cal U}_{L,R}f_{L,R}$ leading to two massless particles, $m_{1,2}=0$, and two massive ones, $m_{3,4}\ne 0$: 
${\cal U}_{R}^\dagger {\cal M}^0_u {\cal U}_{L}={\cal \hat M}^0_u={\tt diag}(0,0,m_3,m_4)$.
The diagram of Fig. \ref{diag} with the exchange
of the $X_\mu$ gauge vector boson, gives rise to the following correction to the mass matrix ${\cal M}^0_u$:
\begin{align}
\delta {\cal M}_u= \frac{g_X^2}{4 \pi^2}\,X_{R}\,{\cal M}^0_u\,{\cal U}_{L}\,  B\,  {\cal U}_{L}^\dagger \,X_{L}\,,
\end{align}
where $B$ 
is a matrix depending on $m_X$, the mass of the gauge vector boson, and the tree-level masses $m_{i}$.
Evaluating  $B$ in dimensional regularization, one gets~\cite{Mohanta:2022seo}:
\begin{align}
{B}_{ij}=(\frac{2}{\epsilon}+1-\gamma+\log 4\pi)~\delta_{ij}-\frac{m_X^2\log m_X^2-m_i^2\log m_i^2}{m_X^2-m_i^2}~
\delta_{ij}\,.
\end{align}
The divergent part of $\delta {\cal M}_u$ is proportional to $X_R\, {\cal M}^0_u\, X_L$ and
the upper-left 3$\times$3 block of $\delta {\cal M}_u$ is a finite function of $g_X$, $m_X$, the charges $X_{L,R}$, and the parameters $\mu_i$, $\mu'_i$ and $M$.
Thus, the two quarks that were massless in the tree approximation acquire a finite non-vanishing mass from 
one-loop corrections. In this particular example, the chiral symmetry acting on the first two quark generations is broken
by the U(1)$_X$ gauge interaction, by assigning different $X_{L,R}$ charges to $u$, $c$ and $t$ quarks.
If $u$, $c$ and $t$ have the same $X_{L,R}$ charges, we can move to an equivalent initial basis where only 
$\mu_3$ and $\mu'_3$ are non-vanishing, a U(2)$^2$ symmetry is manifest, and $\delta {\cal M}_u$ does not
contribute to masses of the first two generations.
If Yukawa interactions are also present,
a further contribution can arise from one-loop diagrams with fermions and scalars in the internal lines.

This simple example illustrates a number of problems common to models in this class. First, the new particles of the theory (fermions, gauge vector bosons and scalar particles) should be sufficiently heavy to evade bounds from current data.
Also, a separation of scales between the first and the second generation should be enforced, for example by
allowing second(first) generation fermion masses to arise at one(two) loop(s). Moreover, aside from the wide range of possible microscopic realizations, a major issue is the large number of independent parameters needed to characterize the new states and their interactions in a realistic model. These parameters must be carefully adjusted to reproduce the observed fermion masses and mixing angles \cite{Nilles:1989pd,Weinberg:2020zba} while remaining consistent with experimental limits on FCNCs. As a result, even though the mass hierarchy among the three generations results from
a chain of loop suppressions, this framework relies on a new dedicated sector whose properties 
pose a new flavor puzzle. Furthermore, it fails to address a fundamental question: why does Nature include exactly three generations of fermions?
\subsection{Infrared fixed points}
\begin{figure}[h!]
	\centering
	\includegraphics[width=.45\textwidth]{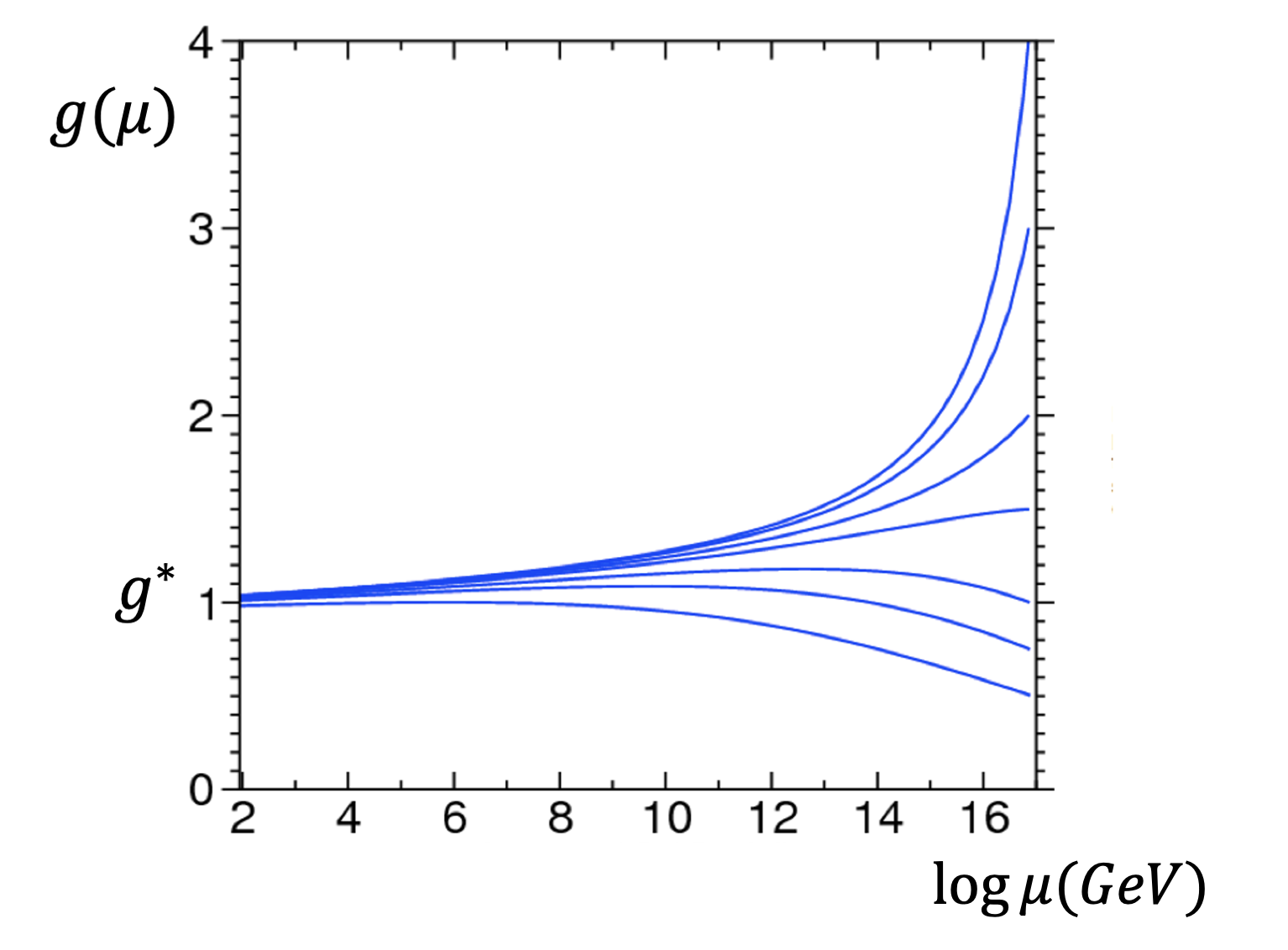}~~~~
	\includegraphics[width=.50\textwidth]{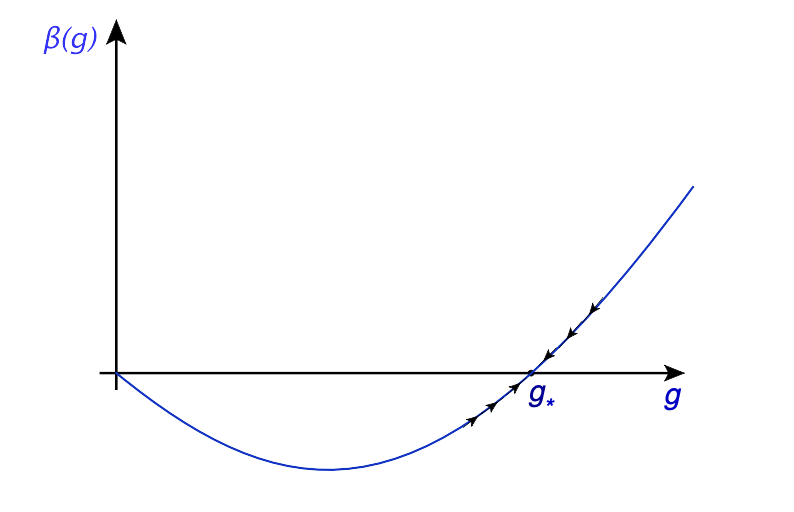}
	\caption{Renormalization group flow of a running coupling $g$ in the presence of an infrared stable fixed point (left). Beta function of a coupling $g$ exhibiting a fixed point at $g=g_*$ (right). Arrows indicate the evolution when the energy is decreased. }
	\label{IRFP}
\end{figure}

There are conditions under which the low-energy parameters of the theory, originating from a set of initial conditions 
assigned at a high energy scale, can completely lose memory of these conditions and be entirely determined by the dynamics of the system. A motivation for the study of this behavior is provided by GUTs, that naturally incorporate the large scale $M_{GUT}\gg v$, where initial conditions are defined. 
An example concerning a single running coupling constant $g(\mu)$ is shown in the left panel of Fig. \ref{IRFP}.
The dynamics predict $g\approx 1$ at low energies for a variety of 
initial conditions set near the GUT scale $M_{GUT}$.
For this to happen, the RGEs of the coupling $g$  should have an infrared stable fixed point, where 
the beta function vanishes and has a positive slope, as shown in the right panel of Fig. \ref{IRFP}. In this case, proceeding from high to low energy scales, the renormalization group flow pushes the coupling to its fixed point value. 
This is indicated by the arrow's flow in the figure.
In general, the renormalized theory depends on a set of running coupling constants
$c_i$, that satisfy RGEs of the type
\begin{align}
\frac{d c_i}{d t}=\beta_i(c)~~~~~~~~~~~~~~t\equiv \frac{1}{16\pi^2}\log\frac{\mu}{M_{GUT}}\,,
\end{align}
where for convenience the initial conditions, at $t=0$, are assigned at the scale $M_{GUT}$.
A fixed point of the theory is a solution $c^*$ of the set of equations $\beta_i(c^*)=0$. It is infrared stable when
$(\partial \beta_i/\partial c_j)$ evaluated at $c=c^*$ is a positive definite matrix. In the SM most of the coupling constants
are small in a wide range of energies. If we neglect all coupling constants but the top Yukawa coupling $Y_t$ and the SU(3) gauge coupling $g_S$, their coupled RGEs in the one-loop approximation have the form
\begin{align}
\label{coupled}
\frac{d Y_t}{d t}=Y_t\,(A\, Y_t^2-D\, g_S^2)\,,~~~~~~~~~~~~~~~~~~~
\frac{d g_S}{d t}=-C\, g_S^3\,.
\end{align}
with $A$, $C$, $D$ constant.
The only fixed point of these equations is the trivial one, where $Y^*_t=g^*_S=0$ corresponding to a non-interacting theory.
However, there is a fixed-point for the ratio $Y_t/g_S$~\cite{Pendleton:1980as}.
Using the variable $u=Y_t^2/g_S^2$ and replacing the scale dependence by the dependence upon the gauge coupling $g_S$, we get
\begin{align}
\label{equ}
C g_S^2\frac{d u}{d g_S^2}=-u\,[A\, u-(D-C)]\,.
\end{align}
In the SM, $A=9/2$, $D=8$ and $C=7$ and this equation has a fixed point at $u^*=2/9$. Since $g_S(\mu)$
grows with decreasing energies, the fixed point is infrared stable. 
The couplings $Y_t$ and $g_S$ are not separately
predicted. When the right-hand side of Eq.~\eqref{equ} vanishes,  $Y_t(\mu)$ and $g_S(\mu)$ are in a fixed ratio which remains constant for all smaller values of $\mu$, independently of their initial condition at high energies.
Using $\alpha_S(m_Z)=g_S^2(m_Z)/4\pi\approx 0.118$ one gets $M_t\approx Y_t(m_Z) v/\sqrt{2}\approx 100$ GeV.
A more precise prediction should include several effects, such as the impact of other parameters -  the gauge couplings $g$ and $g'$ and the Yukawa couplings of the lighter fermions - and higher loop contributions to the RGEs.

More importantly, the fixed point condition refers to
a mathematical limit where $t$ is sent to $-\infty$. In reality $t$ spans a finite range, which does not guarantee that the fixed point relation is realized. Indeed, the full range of $t$ from $M_{GUT}$ down to the electroweak scale, $m_Z$, is only $0.2$.
In practice the ratio $M_{GUT}/m_Z$ is not large enough, so the fixed point solution found above does not
govern the low-energy value of the top-quark Yukawa coupling.

However, it can be shown that when the initial condition satisfies $u(M_{GUT})\gg u^*$, the low-scale values of $u$ are relatively insensitive to the initial value, giving rise to a quasi-fixed point. This can be seen from the solution of Eqs. (\ref{coupled}):
\begin{align}
Y_t^2(t)=\frac{Y_t^2(0) E(t)}{1+2A Y_t^2(0) F(t)}\,,~~~~~~~~~~~E(t)=\left(\frac{g_S^2(t)}{g_S^2(0)}\right)^{D/C}\,,~~~~~~~~~~~~~~F(t)=\int_t^0 E(t') dt'\,.
\end{align}
Formally, when $t$ tends to $-\infty$, $Y^2_t(t)/g_S^2(t)$ tends to the fixed point result $(D-C)/A$. In practice, for sufficiently large initial conditions $Y^2_t(0)$, the solution is well approximated by $Y_t^2(t)=E(t)/2A F(t)$ and exhibits independence from $Y^2_t(0)$. This behavior is called quasi-fixed point~\cite{Hill:1980sq}.
One finds that the quasi-fixed point gives a value for the top
quark Yukawa coupling and the top quark mass approximately twice the true fixed
point value. Moreover, there are also significant corrections due to the electroweak gauge interactions. 
Including these effects the quasi-fixed point predicts a top pole mass $M_t\approx 240$ GeV, too large
to be consistent with the data. In the MSSM, we have $A=6$, $D=16/3$ and $C=3$. The fixed point  is $u^*=7/18$ and the quasi-fixed point solution gives $M_t\approx 200\sin\beta$ GeV, where $\tan\beta=v_u/v_d$ is the ratio between the VEVs $v_{u,d}$ of the two Higgs doublets. 

Fixed points involving lepton mixing angles have also been studied for
neutrino masses of Majorana type, described by the Weinberg operator of Eq. (\ref{weinberg})
originating at some energy scale $\Lambda$ much larger than the electroweak scale $v$.
The elements of the mixing matrix $U_{PMNS}$ vary along the RGE  trajectories
with a speed controlled by
the combination:
\begin{equation}
\frac{\eta}{16\pi^2}y_\tau^2~\frac{m_i+m_j}{m_i-m_j}~\approx~ 6.6\,\eta \,\times 10^{-7}\,\frac{m_i+m_j}{m_i-m_j}\left(\frac{1}{\cos^2\beta}\right)\,,
\end{equation}
where $m_i$ are the neutrino masses, $y_\tau$ is the tau Yukawa coupling and $\eta$ a numerical factor. ($\eta=-3/2$, $y_\tau=\sqrt{2} m_\tau/v$ in the SM; the factor $1/\cos^2\beta$ in parenthesis refer to the MSSM where $\eta=1$, $y_\tau=\sqrt{2} m_\tau/(v \cos\beta)$). Fixed points are reached only if this speed is sufficiently large, which in turn requires a strong degeneracy among neutrino masses. 
In the \CP-conserving case, i.e. neglecting the phases in $U_{PMNS}$, the following fixed point relation has been found
\cite{Chankowski:2001mx}:
\begin{equation}
\sin^2 2\theta^l_{12}=\sin^2\theta^l_{13}\frac{\sin^2 2\theta^l_{23}}{(\sin^2\theta^l_{13}+\sin^2\theta^l_{23}\cos^2\theta^l_{13})^2}~~~.
\label{CPconserving}
\end{equation}
Data rule out this relation by many standard deviations. The allowed 3$\sigma$ ranges for the left-hand and right-hand sides are approximately ($0.75\div 0.92$) and ($0.05\div 0.16$), respectively. 
\paragraph{Summary}
The relative hierarchy among masses of different generations can be explained if lighter masses are radiatively generated
and calculable. This requires an accidental chiral symmetry of the Yukawa sector, removed by other interactions of the theory.
The problem of understanding the Yukawa couplings translates into the problem of understanding the origin and the features of the new sector responsible for the chiral symmetry breaking. 
Fixed-point and quasi-fixed-point structure of the RGEs could help solving the flavor puzzle, by making some
combinations of the low-energy parameters partly insensitive to their initial conditions that, in theories like GUTs,
are set at very high scales. While this program has no success in the SM, it could find realization in its extensions.
\section{Composite Fermions}\label{sec6}
\begin{figure}[h!]
	\centering
	\includegraphics[width=.65\textwidth]{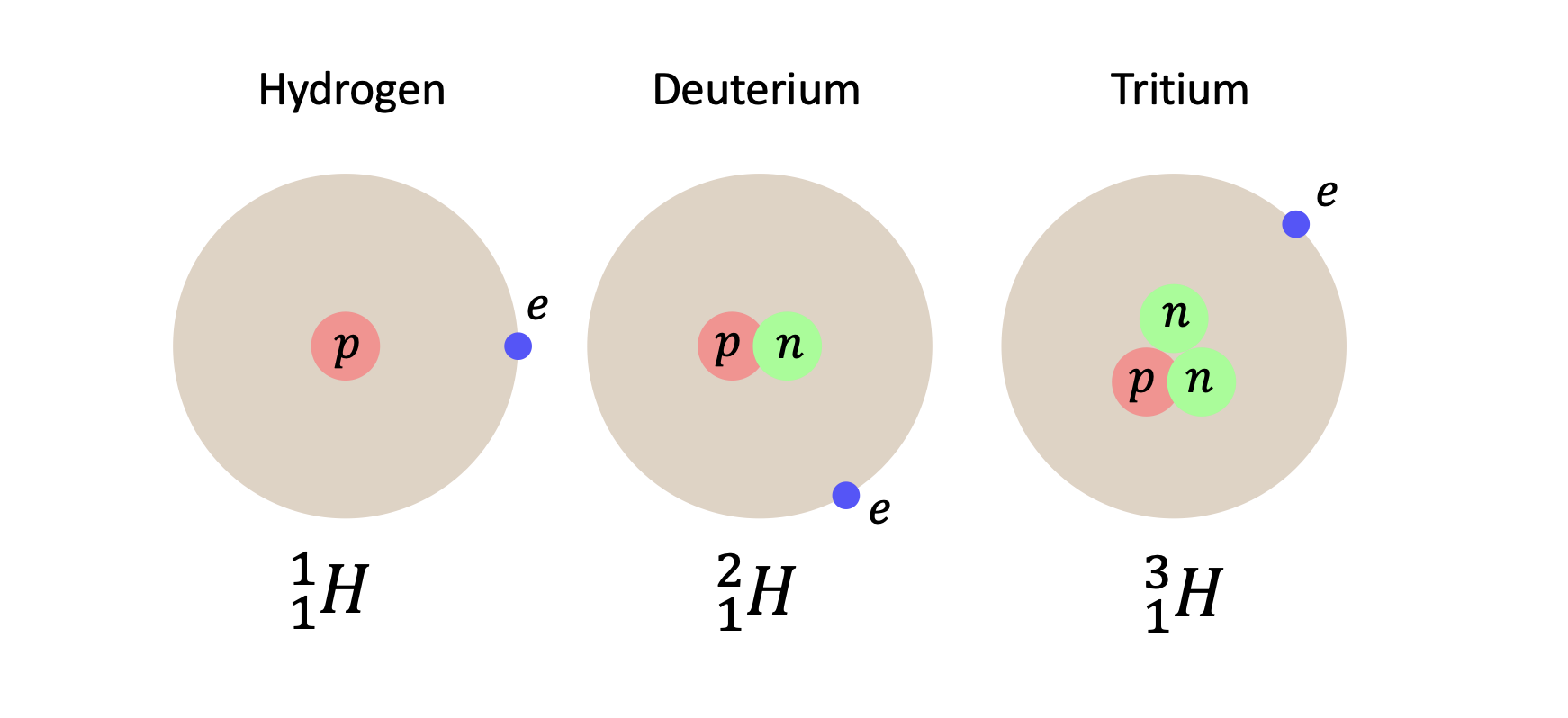}
	\caption{Isotopes of the hydrogen atom.}
	\label{isot}
\end{figure}

A potential explanation for the existence of multiple generations is that SM fermions may not be fundamental, but instead composed of even more basic constituents. 
An analogy can be made between
the SM generations and nuclear isotopes. Consider, for example, the three isotopes of
hydrogen: they have the same chemistry yet have different masses,
a fact simply understood once it is realized that the nucleus is composite, and
that the three isotopes contain a single proton but varying numbers of neutrons, see Fig. \ref{isot}.
So far, three distinct structural layers of matter have been identified: atoms, nucleons, and the fundamental quarks and leptons.
The typical atomic size is approximately $10^{-10}$ m, a scale that is well understood. It corresponds to the inverse of $\alpha M_e$, where $M_e$ is the electron mass, which ultimately determines the characteristic dimensions of atoms and molecules.
The size of nucleons (and atomic nuclei) is around $\Lambda_C^{-1}\approx 10^{-15}$ m, where $\Lambda_C \approx 1$ GeV is the energy scale at which QCD nonperturbative effects and confinement become significant.
Nuclei, which consist of nucleons held together by the strong nuclear force, exhibit sizes comparable to that of individual nucleons. Using the size $r$ of the system to define a compositeness scale $\Lambda=1/r$,
the typical mass of the composite systems known in Nature, such as atoms and nuclei, are greater or approximately equal to the compositeness scale.

The deepest known layer in the structure of matter consists of quarks and leptons. Within the SM, these particles are considered to be pointlike, interacting via gauge boson exchange, with both the fermions and gauge bosons themselves being structureless. Up to now, no experimental evidence contradicts this assumption.
If quarks and leptons do have an internal structure, their size must be smaller than approximately $10^{-20}$ m $\left(\Lambda\sim 10~\text{ TeV}\right)$. 
This stringent upper bound arises from high-precision tests at low energies such as those verifying quantum electrodynamics and from the analysis of high-energy collisions involving quarks and leptons.
For example, analyzing deviations from the electroweak theory parametrized by four-fermion operators of the type
\begin{align}
\frac{4\pi}{\Lambda^2}(\bar q_C \gamma^\mu q_C) (\bar l_{C'}\gamma_\mu l_{C'})\,,~~~~~~~\frac{4\pi}{\Lambda^2}(\bar q_{C} \gamma^\mu q_{C}) (\bar q_{C'} \gamma_\mu q_{C'})\,~~~~~~~~~~~~~(C,C'=L,R)\,,
\end{align}
one gets limits on $\Lambda$ above 10 TeV~\cite{ParticleDataGroup:2024cfk}, for operator combinations involving quarks of the first generation and leptons of the first and second generation.
Thus, in this case the mass of the composite quarks and leptons would be much smaller than
the compositeness scale $\Lambda$.
\subsection{Massless composite fermions}
Having to explain why the fermion masses are significantly smaller than the compositeness scale $\Lambda$, it is natural to seek a framework where these masses vanish~\cite{Peskin:1981ak,Peccei:1982is,Fritzsch:1984ve}. In this approximation, most composite states acquire masses of order $\Lambda$, while quarks and leptons remain massless.
To prevent all composite fermions from acquiring masses of order $\Lambda$, a chiral symmetry is typically introduced. The most direct way to obtain a chiral symmetry for the composite lepton and quark fields is to impose such a symmetry at the constituent level, provided it is preserved by the binding force. A natural candidate for this force is a non-abelian interaction associated with a new gauge group, $G_{MC}$, referred to as metacolor.
The constituents are assumed to carry a metacolor quantum number. The exact number of metacolors and details about the gauge group $G_{MC}$ remain unknown. If metacolor is confining, all physical particles are bound states that are singlets under $G_{MC}$. In a first approximation, quarks and leptons can be metacolorless composite states that remain massless after the theory confines at the scale $\Lambda$. The binding of the constituents due to metacolor is analogous to the color confinement mechanism in QCD.

In such a theory, an approximate chiral symmetry emerges in the presence of massless constituents. This enforces the invariance of the theory under a global group $G_{MF}$, called metaflavor, much like a chiral symmetry arises in QCD when light quark masses are neglected. The expectation is that metaflavor survives at the bound state level and protects composite fermion masses. However, there is no guarantee that this actually happens. QCD provides an example where the chiral symmetry of massless quarks is spontaneously broken, leading to the absence of massless fermions and the appearance of massless pions as Goldstone bosons. Consequently, metachromodynamics must differ significantly from QCD. The key assumption is that metaflavor, or at least a part of it, remains unbroken at the bound state level.
Under this assumption, a powerful tool allows us to extract information about massless composite fermions while bypassing the complexities of a strongly interacting theory: 't Hooft anomaly matching~\cite{tHooft:1979rat}.
\paragraph{'t Hooft anomaly matching}
In general, in the presence of non-abelian metacolor interactions, the metaflavor symmetry is anomalous, and only a part of the metaflavor group remains a symmetry of the quantum theory. In our discussion, $G_{MF}$ refers to this anomaly-free subgroup of the classical global flavor group. Pure global anomalies, arising in the absence of metacolor interactions and related to a triple correlator of global currents, provide crucial insights into the spectrum of massless fermions. Specifically, denoting by $T^A$ the generators of $G_{MF}$ acting on fermion constituents (in a basis where all fermions have the same chirality), the anomaly is proportional to $\text{tr}(T^A\{T^B,T^C\})$.

The metaflavor group can undergo spontaneous symmetry breaking, as in the case of chiral symmetry in QCD. However, suppose it remains unbroken along the renormalization flow from high energies, where the theory is asymptotically free, down to low energies, where the theory confines. 't Hooft's remarkable observation is that, in this case, the global anomalies must remain unchanged throughout the entire flow. Consequently, if the anomaly $\text{tr}(T^A\{T^B,T^C\})$ is nonzero at the constituent level, there must exist a set of massless composite fermions that reproduce the same anomaly in the confined theory. Despite the complexity of strongly interacting dynamics and the intricacies of infrared physics, the easily computable coefficients $\text{tr}(T^A\{T^B,T^C\})$ impose constraints on the spectrum of massless composite fermionic bound states.

As an example~\cite{Dimopoulos:1980hn,Peskin:1981ak}, consider a gauge theory where $G_{MC} = \SU5$ with two left-handed constituent fermions, $\psi_{10}^{ab}$ and $\psi_{\bar{5}a}$ ($a,b=1,...,5$), transforming under the $\rep{10}$ and $\crep{5}$ representations, respectively. No mass term is allowed for the constituents. At the classical level, the theory exhibits a global invariance under the group $\U1_{10}\x\U1_{\bar 5}$:
\begin{align}
\psi_{10}^{ab} \to e^{\I \alpha_{10}}\psi_{10}^{ab}\,,
\qquad
\psi_{\bar{5}a} \to e^{\I \alpha_{\bar{5}}}\psi_{\bar{5}a}\,.
\end{align}

In the presence of gauge interactions from $G_{MC} = \text{SU}(5)$, this global symmetry is broken by quantum effects down to a U(1) subgroup of the above transformations, obtained by setting $\alpha_{\bar{5}}=3\alpha$ and $\alpha_{10}=-\alpha$. This remaining symmetry is the metaflavor group of the theory, $G_{MF} = \U1$. At the constituent level, the pure global anomaly is given by the sum of the third power of the charges over all fermion components:
\begin{align}
5 \times 3^3 + 10 \times (-1)^3 = 125\,.
\end{align}
A set of composite fermions matching this anomaly must exist. A candidate massless composite fermion is the color-singlet combination
\begin{align}
\psi \sim \psi_{10}^{ab}\psi_{\bar{5}a}\psi_{\bar{5}b}\,,
\end{align}
which carries a metaflavor charge of $-1+3+3=5$, whose cube indeed reproduces 125.
\paragraph{Supersymmetric composite models}
Leptons and quarks can be bound states of fermionic constituents, as in the previous example, or of fermionic and bosonic states. Bound states composed of fermions and bosons naturally arise in a SUSY theory. In the limit of unbroken SUSY, composite fermions can remain massless if their bosonic partners also remain massless.
Various SUSY Yang-Mills theories exhibit dualities that provide a natural setting for composite fermions, ensuring 't Hooft anomaly matching~\cite{Seiberg:1994bz,Seiberg:1994pq}. One illustrative, though unrealistic, example is an Sp(6) gauge theory with six fundamentals $Q^i_a$ ($i=1,\ldots,6$) and an antisymmetric tensor $A_{ab}$ as chiral supermultiplets~\cite{Kaplan:1997tu}. The theory has an \SU6\x\U1 global symmetry and an R-symmetry. It confines without breaking chiral symmetry, yielding three generations of metacolor singlets
(Sp(6) indices are contracted with the appropriate metric):
\begin{align}
M_0^{ij} \sim Q^i Q^j\,,~~~~~~~~M_1^{ij}  \sim Q^i A Q^i\,,~~~~~~~~M_2^{ij}  \sim Q^i A^2 Q^j\,,
\end{align} 
where only the antisymmetric combination of the indices $ij$ survives, leading to 15 flavor states per generation.
In addition, there are two neutral exotic states
\begin{align}
T_2 \sim {\rm Tr} A^2\,,~~~~~~~~~~T_3 \sim {\rm Tr} A^3\,.
\end{align}
With this set of light multiplets, the 't Hooft anomaly matching is satisfied. 
The model realizes the isotope paradigm, with the $QQ$ and $A$ fields playing the roles of the proton and neutron respectively. The quantum numbers of constituents and composite fields are shown in Table~\ref{tab:composites}.
\begin{table}[h!]
\begin{center}
	\TBL{\caption{\label{tab:composites}Representations of constituents and composites under metacolor and metaflavor groups.
	Representations are denoted by Young tableaux:
$\Yfund$ is the defining representation, while $\Yasymm$ denotes an
antisymmetric tensor.}}{%
		\begin{tabular*}{0.50\columnwidth}{@{\extracolsep\fill}ccccc@{}}
			\toprule
			   & Sp(6) & SU(6) & U(1) & U(1)$_R$ \\
			\colrule
		  $A$ & \Yasymm   & $1$     & $-3$   & 0 \\[0.1 cm]
    $Q$ & \Yfund    & \Yfund& $+2$& $\frac{1}{3}$ \\	
    \colrule
  $T_m$    & $1$ & $1$       & $-3m$  & 0 \\[0.1 cm]
  $M_n$   &  $1$ & \Yasymm & $4 -3 n$ & $\frac{2}{3}$\\
			\botrule
	\end{tabular*}}{}
	\end{center}
\end{table}
This model has a number of desirable features. If a subgroup of \SU6 is gauged, then $M_{0,1,2}$ realize a replication of
families with the same gauge quantum numbers. Furthermore, the global \U1 symmetry distinguishes among the three
generations and plays the role of a horizontal or family symmetry, see Section~\ref{sec7}.
The model is not viable, since a SM family cannot fit in the antisymmetric $\rep{15}$ representation of \SU6.
In addition to that, electroweak symmetry, SUSY and the family symmetry should be broken in a realistic model.
At the same time 
the chiral symmetry, tailored to keep quark and lepton massless, should be broken by a carefully chosen sector
and small fermion masses, compared to the compositeness scale $\Lambda$, should appear.
In general, all these effects depend on both the electroweak scale $v$ and the scale $\Lambda$, and their inclusion entails several difficulties. 
The result must ensure the suppression of FCNCs, protected by the GIM mechanism in the SM, and the suppression of processes involving baryon number violation. The amplitudes of such processes are typically suppressed by inverse powers of the scale $\Lambda$, which would need to be pushed many orders of magnitude above the energies currently explored. The construction of realistic models has proved difficult and no ``standard'' composite model has emerged in the literature. So far, attempts in this direction have not advanced the calculability of physical parameters.
\subsection{Partial compositeness}\label{sec6dot2}
The generation of fermion masses is intimately tied to the mechanism of electroweak symmetry breaking and is therefore directly influenced by the nature of the Higgs boson. In scenarios where quarks and leptons are composite states, the Higgs boson may also be composite. To address the hierarchy problem - the large disparity between the electroweak and Planck scales - it is desirable to introduce a compositeness scale $\Lambda$ around 1 TeV.
However, a compositeness scale as low as 1 TeV is insufficient to evade stringent experimental bounds on the compositeness of the first-generation quarks and leptons. To alleviate these constraints, one can assume that SM fermions are not fully composite, but rather admixtures of elementary and composite states~\cite{Kaplan:1991dc}. In this framework, a strongly interacting sector at the TeV scale gives rise to heavy composite fermions that couple with order-one strength to the Higgs boson. SM fermions then acquire masses through their mixing with these bound states. The larger the mixing between an SM fermion and its composite partner, the higher its degree of compositeness and, consequently, its mass. Heavier fermions are thus predominantly composite, while lighter ones remain mostly elementary.

As a toy model illustrating this idea, consider a scenario for quarks in which, for each SM fermion, the composite sector contains a pair of heavy fermions that allow for a Dirac mass term of order $\Lambda$ as well as a mixing term with SM fields~\cite{Contino:2006nn}:
\begin{align}
\mathcal{L} &= -\bar{Q}_R M_Q Q_L - \bar{D}_R M_D D_L - \bar{U}_R M_U U_L + \text{h.c.} \\
&\quad - \bar{Q}_R \tilde{Y}_D \tilde{\varphi}^\dagger D_L - \bar{D}_R Y_D \tilde{\varphi} Q_L - \bar{Q}_R \tilde{Y}_U \varphi^\dagger U_L - \bar{U}_R Y_U \varphi Q_L + \text{h.c.} \\
&\quad - \bar{Q}_R \Delta_Q q_L - \bar{d}_R \Delta_D D_L - \bar{u}_R \Delta_U U_L + \text{h.c.}
\end{align}
The first line contains Dirac mass terms for the composite fermions, with $M_{Q,D,U} \sim \Lambda$. The second line describes Yukawa interactions within the composite sector, characterized by strong couplings $Y_{U,D}, \tilde{Y}_{U,D} \gtrsim 1$. The third line introduces mixing terms between the elementary and composite sectors, parameterized by the matrices $\Delta_{Q,D,U}$.
By integrating out the heavy vector-like quarks in the static limit, we obtain an effective Lagrangian of the form:
\begin{align}
\mathcal{L}_{\text{eff}} = -\bar{d}_R\, X_D \left( Y_D\, \tilde{\varphi} + \dots \right) X_Q\, q_L - \bar{u}_R\, X_U \left( Y_U\, \varphi + \dots \right) X_Q\, q_L + \text{h.c.},
\end{align}
where
\begin{align}
X_D \equiv \Delta_D M_D^{-1}, \qquad
X_U \equiv \Delta_U M_U^{-1}, \qquad
X_Q \equiv M_Q^{-1} \Delta_Q,
\end{align}
and the dots represent terms with three or more powers of the Higgs field.
Neglecting terms of order $v^2/\Lambda^2$, the effective Yukawa matrices for SM quarks become:
\begin{align}
\label{partcomp}
Y_d = X_D Y_D X_Q, \qquad\qquad Y_u = X_U Y_U X_Q.
\end{align}
If the entries of $Y_{U,D}$ are of order one, then the observed hierarchies in quark masses and mixing angles arise from the hierarchical structure of the matrices $X_{D,U,Q}$. In this picture, light fermions such as the up and down quarks are mostly elementary, i.e., $(X_{D,U,Q})_{11}, (X_{D,U,Q})_{22} \ll 1$, while heavy fermions like the top quark are predominantly composite, with $(X_{U,Q})_{33} \approx 1$.
\begin{figure}[h!]
	\centering
	\includegraphics[width=.35\textwidth]{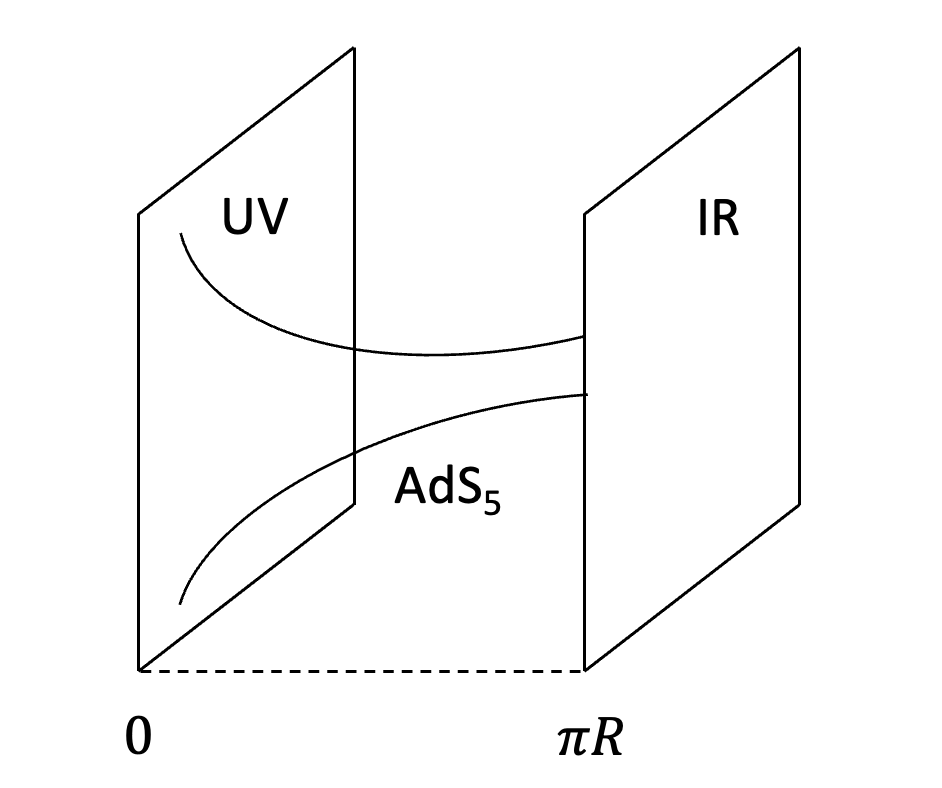}~~~\includegraphics[width=.55\textwidth]{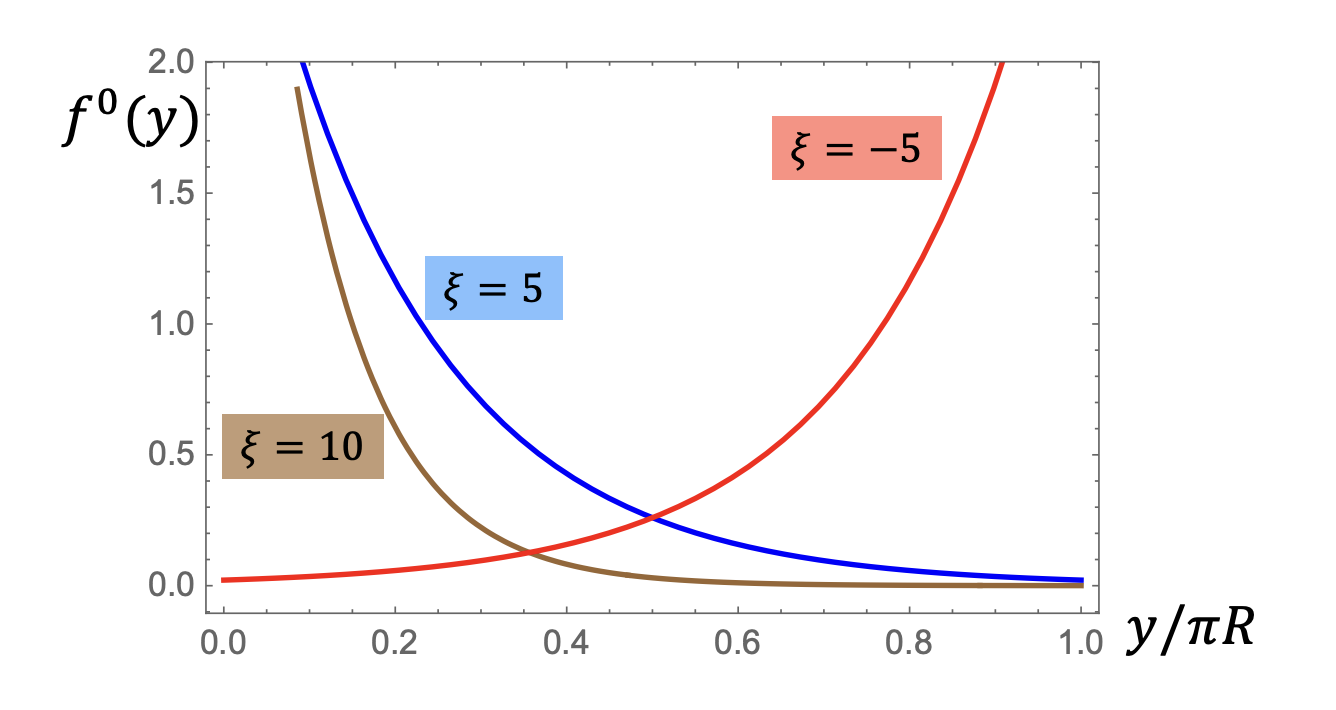}
	\caption{Left: pictorial view of spacetime in the RS model. Right: profiles of a few fermion zero modes.}
	\label{sec2:fig1}
\end{figure}
\paragraph{Dual description}
This dynamic allows for a dual description, based on the AdS/CFT correspondence~\cite{Maldacena:1997re}. This is a conjectured duality between a gravitational theory in a $(d+1)$-dimensional AdS space - the bulk - 
and a conformal field theory living on its $d$-dimensional boundary. It provides a realization of the holographic principle and an example of strong/weak duality. The higher-dimensional gravity theory is equivalent to a lower-dimensional quantum field theory without gravity. Strong coupling in the CFT corresponds to weak gravity in AdS. The simplest example is provided by the RS model~\cite{Randall:1999vf}, where the 4-dimensional (4D) space-time is extended by adding a fifth dimension describing an orbifold $S/Z_2$. On the circle $S$, the points of coordinate $y$ and $-y$ are identified and the extra dimension has the geometry of an interval, $0\le y\le \pi R$. At the endpoints of the interval
two 4D space-times - the branes - are defined, the UV brane at $y=0$ and the IR brane at $y=\pi R$, see Fig. \ref{sec2:fig1}. If these branes carry some energy density, by solving the Einstein equations one finds an AdS non-factorizable metric
\begin{equation}
\label{ads}
  ds^2 = e^{-2k |y| } \eta_{\mu\nu} dx^\mu dx^\nu - dy^2\,,
\end{equation}
where $\eta_{\mu\nu}={\tt diag}(+---)$ is the Minkowski metric and $k$ is the curvature scale of the AdS space. The RS model was proposed to solve the hierarchy problem by localizing the Higgs field near the IR brane.
Any mass parameter $M_0$ in the fundamental theory gives rise to an effective mass parameter $M = e^{-k R \pi} M_0$ governing the physical properties of particles on the brane at $y=\pi R$. If $M_0=M_{Pl}$ and $k R\approx 11$, the mass $M$ felt on the IR brane is about 1 TeV.

In the dual CFT picture, the Higgs field localized near the IR brane is a composite scalar.
Depending on their localization in the fifth dimension, bulk fermions correspond to an admixture between elementary fermions and composite CFT operators. Bulk fermions localized near the UV/IR brane are mostly elementary/composite.
The Yukawa couplings in four dimensions are determined by wavefunction overlaps in five dimensions. Light fermions (like the first-generation quarks and leptons) are mostly elementary, their overlap with the Higgs is small, leading to small masses. In contrast, heavy fermions (like the top quark) are more composite and thus have a larger overlap with the Higgs, generating larger masses. It is instructive to derive the Yukawa couplings in the limit of infinite AdS radius, $k\to 0$, corresponding to a 5D flat space-time.
A 5D spinor $\Psi(x,y)$ can be decomposed in two 4D spinors carrying opposite chiralities
\begin{align}
\Psi=
\left(
\begin{array}{c}
\Psi_L\\
\Psi_R
\end{array}
\right)\,.
\end{align}
Consistency with the orbifold construction requires the 4D spinors to have opposite $\Z2$ parities: if $\Psi_L(x,-y)=\pm \Psi_L(x,y)$, then $\Psi_R(x,-y)=\mp \Psi_R(x,y)$.
The Lagrangian of a massive 5D spinor reads:
\begin{align}
{\cal L}=&~i\overline{\Psi}\Gamma^M D_M \Psi+M_0 \epsilon(y)\overline{\Psi} \Psi\nn\\
=&~i\overline{\Psi}\gamma^\mu \partial_\mu \Psi-\overline{\Psi}\gamma_5 \partial_y \Psi+M_0 \epsilon(y)\overline{\Psi} \Psi+...
\end{align}
where $D_M$ is the gauge covariant derivative and $\epsilon(y)$ is the sign function, required to make ${\cal L}$ an even function of $y$. Only the even component of $\Psi(x,y)$ develops a zero mode.
Choosing, for instance, $\Psi_L$ even and $\Psi_R$ odd, the equation satisfied by the zero mode of $\Psi_L$ is \begin{align}
\partial_y \Psi_L^0+M_L\epsilon(y)\Psi_L^0=0\,,~~~~~~~~~~~~(M_0=M_L)\,,
\end{align}
whose solution has an exponential dependence on $y$. Using the normalization $1/\pi R\int_0^{\pi R} dy |f^0_L(y)|^2=1$, one finds
\begin{align}
\label{solflat}
\Psi^0_L(x,y)=\frac{1}{\sqrt{\pi R}}\psi_L(x)\,f^0_L(y)\,,~~~~~~~~~~~~~~~f^0_L(y)=\sqrt{\frac{2\xi_L}{1-e^{-2\xi_L}}}\,e^{\dd -\xi_L |y|/ \pi R}\,,~~~~~~~~~~~~~~\xi_L\equiv M_L\pi R\,.
\end{align}
The zero mode is localized near $y=0\,(\pi R)$ for $\xi_L>0\,(<0)$, see Fig. \ref{sec2:fig1}. Choosing $\Psi_L$ odd and $\Psi_R$ even
and starting from a 5D mass term with the opposite sign, we find a formally identical solution for the zero mode of $\Psi_R$. If the Higgs field is strictly localized at one of the two branes, for instance the one at $y=\pi R$, the 5D Yukawa interaction among the Higgs doublet and the fermion zero modes $\Psi^0_{L,R}(x,y)$ translates into the 4D Yukawa coupling
\begin{align}
\label{4Yuk}
\frac{\lambda}{\Lambda\pi R} \int_0^{\pi R} f^0_R(y)\,f^0_L(y)\,\delta(y-\pi R)\, dy=\sqrt{\frac{2\xi_R}{e^{+2\xi_R}-1}}\,Y\,\sqrt{\frac{2\xi_L}{e^{+2\xi_L}-1}}\,,~~~~~~~~~~~~~~~Y\equiv \frac{\lambda}{\Lambda\pi R}\,.
\end{align}
Here $\Lambda$ is the typical scale of the 5D theory, and $\lambda$ is a dimensionless coupling constant.

In the AdS metric of Eq.~(\ref{ads}), 5D fermions with bulk masses $(c_{L,R} \, k)$ give rise to zero modes that reproduce the same result as in Eq.~(\ref{4Yuk}), upon the replacement $M_{L,R} \rightarrow (c_{L,R} - 1/2)\,k$~\cite{Gherghetta:2000qt,Gherghetta:2010cj}. 
Consider three generations of 5D quarks $U_i, D_i$, and $Q_i$, with bulk masses $(c_{U_i}k)$, $(c_{D_i}k)$, and $(c_{Q_i}k)$, respectively, coupled to a Higgs localized on the IR brane at $y = \pi R$, with generic couplings $(Y_{U,D})_{ij}$. For the corresponding 4D fields $u_{Ri}$, $d_{Ri}$, and $q_{Li}$, associated with the zero modes, the resulting Yukawa couplings take the same form as in Eq.~(\ref{partcomp}), with:
\begin{align}
(X_A)_{ii} = \sqrt{\frac{2\xi_{A_i}}{e^{+2\xi_{A_i}} - 1}} \approx
\begin{cases}
\sqrt{2\xi_{A_i}} \, e^{-\xi_{A_i}}, & \xi_{A_i} > 0 \\
\sqrt{-2\xi_{A_i}}, & \xi_{A_i} < 0
\end{cases}
\quad \text{for } A = U, D, Q\,.
\end{align}
This setup mirrors the framework of partial compositeness, with the key difference that the structure of the Yukawa couplings emerges geometrically from the compact extra dimension. If the couplings $Y_{U,D}$ on the IR brane are anarchic (i.e., all entries of the same order), hierarchical masses and mixing angles can be generated via the matrices $X_{D,U,Q}$. Due to the exponential sensitivity of $X_{D,U,Q}$ to the bulk parameters $c_{U_i}$, $c_{D_i}$, and $c_{Q_i}$, small mass ratios and mixing angles can be produced with only modest hierarchies in the underlying parameters~\cite{Gherghetta:2000qt,Huber:2003tu}.

\paragraph{Flavor Constraints and Higher-Dimensional Operators}

Without further assumptions, this framework permits higher-dimensional operators on the IR brane - such as those responsible for proton decay and neutrino masses - suppressed only by $M = e^{-k\pi R} M_0$, typically close to the TeV scale. This suppression is generally insufficient to comply with current experimental bounds. To resolve this issue, additional mechanisms are needed. For instance, by imposing lepton number conservation, Dirac neutrino masses can be accommodated~\cite{Grossman:1999ra}. Dirac neutrino masses can naturally emerge from geometric effects, even if lepton number is explicitly broken in the ultraviolet theory~ \cite{Gherghetta:2003hf}.
Both in partial compositeness and in its 5D dual, we expect new states with flavor quantum numbers and masses near the compositeness scale $\Lambda$. Assuming anarchic Yukawa couplings $Y_{U,D}$, these states represent new potential sources of FCNCs and \CP violation, leading to strong bounds on $\Lambda$. A particularly stringent constraint arises from the effective operator:
\begin{align}
\frac{1}{\Lambda^2}(\bar{q}_L X_Q^\dagger \gamma^\mu X_Q q_L)(\bar{d}_R X_D \gamma_\mu X_D^\dagger d_R) 
\approx \frac{1}{\Lambda^2 \langle Y_D^2 \rangle} \frac{2 m_d m_s}{v^2} (\bar{s}_L d_R)(\bar{s}_R d_L) + \dots,
\label{estimate}
\end{align}
where $\langle Y_D^2 \rangle$ denotes an average $\mathcal{O}(1)$ coupling. The contribution of this operator to the \CP-violating parameter $\epsilon_K$ is amplified both by hadronic matrix elements and QCD corrections. Assuming a generic $\mathcal{O}(1)$ \CP-violating phase, consistency with the observed value of $\epsilon_K$ imposes the bound:
\begin{align}
\langle Y_D \rangle \, \Lambda > 20~\text{TeV}.
\end{align}
A similar analysis applies to the charged lepton sector.
In analogy with the quark sector, the matrix of charged lepton Yukawa couplings $Y_e$ can be decomposed as $Y_e=X_E Y_E X_L$ and one can construct the dipole operator:
\begin{align}
\frac{e}{\Lambda^2} \bar{e}_R \sigma_{\mu\nu} F^{\mu\nu} (X_E Y_E Y_E^\dagger Y_E X_L) \varphi^\dagger \ell_L.
\end{align}
In general, the matrices $X_E Y_E X_L$ and $X_E Y_E Y_E^\dagger Y_E X_L$ are not diagonal in the same basis, leading to predictions for radiative decays such as $\mu \to e\gamma$ and $\tau \to \mu\gamma$. Compatibility with the current experimental limit, $BR(\mu \to e \gamma) < 3.1 \times 10^{-13}$~\cite{MEGII:2023ltw}, requires $\Lambda$ to be well above $10$ TeV.
These considerations suggest that a fully anarchic structure for $Y_{U,D,E}$ may not be viable if flavored new physics exists at the TeV scale.

\paragraph{Summary}
Composite fermions with masses much smaller than their inverse radii can arise in confining theories endowed with approximate chiral symmetries. However, building realistic models along these lines has proven difficult, and no standard framework has emerged in the literature. Furthermore, to satisfy experimental constraints, the compositeness scale must be pushed above 10 TeV.
An alternative is that SM fermions are only partially composite, with their degree of compositeness determined by the mixing between elementary and composite sectors. The 5D warped-space dual of this idea provides a geometric interpretation of flavor hierarchies and a concrete realization of composite Higgs models.
Unfortunately, in its minimal version, this framework does not adequately suppress FCNCs when new physics appears near the TeV scale. Additionally, existing models often involve a large number of free parameters, limiting their predictive power and calculability.
\section{Flavor Symmetries}\label{sec7}

A common feature of many of the above frameworks is the presence of a symmetry acting in generation space - a flavor symmetry - that emerges when the masses of the lighter generations (or the entire electroweak scale) are neglected. It remains unclear whether this symmetry reflects a fundamental property of Nature or simply arises from the way we approximate and model the problem in successive steps.
In the case of fermion localization in a five-dimensional space, such a symmetry seems to be an artifact: the couplings between fermions and the Higgs on the IR brane are generic, and the observed hierarchies in masses and mixing angles stem from the geometry of the extra dimension. 
Conversely, an example of how symmetries played a key role in classifying particle properties and developing a fundamental theory is the Eightfold Way, where hadrons were organized into \SU3 multiplets of nearly degenerate mass.
Even if one believes flavor symmetries hold the key to the flavor puzzle, Nature does not seem to provide clear guidance on the choice of the flavor group $G_f$ or the representations under which quarks and leptons transform. Many plausible options exist, yet no single choice stands out as clearly preferred.

\subsection{The Froggatt-Nielsen Model}

In a seminal paper~\cite{Froggatt:1978nt}, Froggatt and Nielsen observed that ratios of charged fermion masses and CKM matrix elements can be expressed as powers of the Cabibbo angle. Using $\lambda = 0.22$, we have
\begin{align}
\frac{m_e}{m_\tau} \approx \lambda^{5.4}, && \frac{m_d}{m_b} \approx \lambda^{4.3}, && \frac{m_u}{m_t} \approx \lambda^{7.4}, \\
\frac{m_\mu}{m_\tau} \approx \lambda^{1.9}, && \frac{m_s}{m_b} \approx \lambda^{2.3}, && \frac{m_c}{m_t} \approx \lambda^{3.6},
\end{align}
where all masses are evaluated at the scale $m_Z$. 
Similarly, for the elements of the CKM mixing matrix, one finds
\begin{align}
|V_{ud}|\approx 1\,,\qquad
|V_{us}|\approx\lambda\,,\qquad
|V_{cb}|\approx\lambda^2\,,\qquad
|V_{ub}|\approx\lambda^4\div\lambda^3\,.
\end{align}
Froggatt and Nielsen suggested these small quantities can be viewed as powers of a symmetry breaking parameter.
The theory is invariant under an abelian flavor symmetry, $G_f=~$U(1)$_{F}$.
A scalar field $\varphi$, carrying by convention a negative unit of the abelian charge $F$, develops a VEV that - evaluated in units of a fundamental scale $\Lambda_f$ - reads
\begin{align}
\lambda=\langle \varphi\rangle/\Lambda_f<1~~~~~~~~~~~~~F(\varphi)=-1 ~~~.
\end{align}
The flavor symmetry is broken. The higher the $F$ charge of a fermion bilinear, the more suppressed is the corresponding mass term. If electroweak doublets and (conjugate) singlets $(A_i=q_{Li},\bar u_{Ri},\bar d_{Ri})$ all carry positive charges $F(A_i)\ge 0$, the quark Yukawa couplings $Y_{u,d}$  read
\begin{align}
\label{FN}
Y_d=X_D Y_D X_Q\,,\qquad
Y_u=X_U Y_U X_Q\,.
\end{align}
Here $Y_{U,D}$ are complex matrices whose entries cannot be determined by the ${\rm U(1)}_{F}$ symmetry, 
while $X_{D,U,Q}$ are real diagonal matrices, depending on powers of $\lambda$ specified by the charges $F(A_i)$:
\begin{align}
\label{XA}
X_A=
\left(
\begin{array}{ccc}
\lambda^{F(A_1)}&0&0\\
0&\lambda^{F(A_2)} &0\\
0&0&\lambda^{F(A_3)}
\end{array}
\right)~~~~~~~~~~~~~~~~~~~~~~~(A=U,D,Q)\,.
\end{align}
In this picture, the Yukawa interaction comes from higher-dimensional operators suppressed by inverse powers
of the scale $\Lambda_f$. The result of Eq.~(\ref{FN}) is identical to the one in Eq.~(\ref{partcomp}), obtained within partial compositeness or fermion localization in 5D, with the difference that in the FN models there is no preferred value for the scale $\Lambda_f$, since mass ratios and mixing angles depend on $\Lambda_f$ only through the combination $\langle \varphi\rangle/\Lambda_f$. Thus $\Lambda_f$ can be much larger than the electroweak scale, and the UV completions of
the model can easily evade bounds from FCNC. 

If the unknown matrices $Y_{U,D}$ have all entries of order one, the small quark mass ratios and quark mixing angles originate from the hierarchical structure of the matrices $X_A$. By taking $F(q_{L1})>F(q_{L2})>F(q_{L3})\ge 0$ we get 
\begin{align}
(V_{u,d})_{ij}\approx \lambda^{F({q_{Li}})-F({q_{Lj}})}~~~~~~~~~~~~~~~(i<j)\,,
\end{align}
for the matrices $V_{u,d}$ diagonalizing $(m_{u,d}^\dagger m_{u,d})$ and defining the CKM mixing matrix as $V_{CKM}=V_u^\dagger V_d$. 
Independently from the specific charge choice, this framework predicts
\begin{align}
V_{ud}\approx V_{cs}\approx V_{tb}\approx O(1)\,,~~~~~~~~~~~~~~~~~~~~V_{ub}\approx V_{td}\approx V_{us}\times V_{cb}\,,
\label{Vij}
\end{align}
the last equality being correct within a factor of two. With $\lambda\approx 0.2$, the correct order of magnitude of the $V_{CKM}$ matrix elements can be reproduced by choosing
$F(q_L)=(3,2,0)$. Quark mass ratios are approximately reproduced by the choice
\begin{align}
F(q_L)=(3,2,0)~~~~~~~~~~F(\bar u_R)=(4,2,0)~~~~~~~~~~F(\bar d_R)=(1+r,r,r)~~~,
\label{qcharges}
\end{align}
$r$ being a non-negative integer. If there is only one Higgs doublet, $r$ should be close to 2 to match the ratio $m_t/m_b$. If two Higgs doublets are present, other choices are possible by varying $\tan\beta=v_u/v_d$. 

In the lepton sector we have no evidence for strong hierarchies among mixing angles or neutrino masses,
except for the $|U_{e3}|$ element of the mixing matrix $U_{PMNS}$, which is of order $\lambda$, and the ratio between the solar and the atmospheric neutrino mass-squared differences $\delta m^2/|\Delta m^2|$, of order $\lambda^2$. Hierarchy shows up mainly at the level of charged lepton masses. These features can be
reproduced by choosing FN charges
$F({\bar e_{R1}})> F({\bar e_{R2}})> F({\bar e_{R3}})$ and $F({l_{L1}})\approx F({l_{L2}})\approx F({l_{L3}})$.
For example
\begin{align}
F(\bar e_R)=(4,2,0)~~~~~~~~~~F(l_L)=(s+t,s,s)~~~~~~~~~~(s\ge0,t=0,1)~~~.
\label{lcharges}
\end{align}
The Yukawa couplings of the lepton sector can be built from matrices $X_{E,L}$, analogous of the matrices $X_{U,D,Q}$ of the quark sector, Eq.~(\ref{XA}). If neutrinos are of Majorana type, and $W^\nu_{ij}$ is the set of couplings in the Weinberg operator of Eq.~(\ref{weinberg}), we get
\begin{align}
\label{Yleptons}
Y_e=X_E\, Y_E\, X_L~~~~~~~~~~~~~~~~~~~~W^\nu=X_L Y_W X_L\,,
\end{align}
where the entries of the matrices $Y_{E,W}$ are free parameters, assumed to be of order one.
The seesaw mechanism can help explaining the smallness of $\delta m^2/|\Delta m^2|$ that, within this simple framework, is accidental. A small $|U_{e3}|$ can be obtained by choosing $t=1$. Otherwise, if one regards
$|U_{e3}|$ as a quantity of order one, also the choice $t=0$ is acceptable, leading to a completely anarchical neutrino sector~\cite{Hall:1999sn,Altarelli:2012ia}.

This simple flavor symmetry can also be implemented in a SU(5) GUT, where quarks and leptons are hosted in the same multiplet of the gauge group. The SU(5) gauge symmetry requires particles within the same multiplet to have equal $F$ charges, resulting in
\begin{align}
X_Q=X_U=X_E=X_{10}\,,~~~~~~~X_L=X_D=X_{\overline{5}}\,,
\label{su5}
\end{align}
for the $\rep{10} = (q_L, \bar{u}_R, \bar{e}_R)$ and $\crep{5} = (l_L, \bar{d}_R)$ representations. The previous options, Eqs.~(\ref{qcharges}) and (\ref{lcharges}), come very close to this requirement if we choose $r=s$ and $t=1$. If we accept a couple of tunings in the unknown $O(1)$ parameters $Y_{U,D}$,
we can force the equality (\ref{su5}) and still have a decent description of both the quark and lepton mass spectrum.
The large lepton mixing corresponds to a large mixing among $\bar d_R$ quarks, unobservable in SM weak interactions.

\subsection{Variants and alternatives}\label{sec:variants}
The abelian $\U1_{F}$ proposed by FN represents the prototype of flavor symmetry. It successfully explains the  observed pattern of masses and mixing angles, with input parameters of the same order of magnitude.
However, abelian flavor symmetries are too weak and fail in making precise predictions due to the large number of unknown parameters in $Y_{U,D}$. No conclusive test, matching the present level of experimental accuracy, can be planned. On the other hand, the main ingredients of the FN idea can be varied to build a large variety of models, in the hope of increasing predictability.

As a first step, one chooses the flavor group $G_f$, and then assigns SM fermions to representations of $G_f$.
For example, in the quark sector, the multiplets $A=(q_{L},\bar u_{R},\bar d_{R})$ are assigned to - possibly reducible - representations $\rho_A$ of $G_f$. To build a $G_f$ invariant Lagrangian, the Yukawa couplings $Y_u$ and $Y_d$ should transform according to representations contained in the products $\rho_{u_{R}}\times \bar\rho_{q_{L}}$ and $\rho_{d_{R}}\times \bar\rho_{q_{L}}$, respectively, and eventually
one builds $Y_u$ and $Y_d$ out of a set of scalar fields $\tau_\alpha$ transforming in that way.
The fields $\tau_\alpha$ play the key role of symmetry breaking sector. In the first place, Yukawa couplings are field-dependent quantities, which become numbers once the fields $\tau_\alpha$ are set to their VEVs.
A generic Yukawa coupling $Y$ can be expanded in powers of $\tau_\alpha$ and $\bar\tau_{\bar\alpha}$ - expressed in units of some fundamental scale $\Lambda_f$ - as
\begin{align}
\label{expY}
Y(\tau,\bar\tau)=Y^0+Y^\alpha\tau_\alpha+Y^{\bar\alpha}\bar\tau_{\bar\alpha}+Y^{\alpha\beta}\tau_\alpha\tau_\beta+...\,,
\end{align}
where $Y^0$, $Y^\alpha$, $Y^{\bar\alpha}$, $Y^{\alpha\beta}$,... are coefficients constrained by group theory.
In general, each coefficient depends on a set of input parameters. For example, if the product $(\bar u_R\, \tau\, q_L )$
contains $n$ independent invariants under $G_f$, the coefficient $Y^\alpha$ depends on $n$ unconstrained parameters.
The first term of the expansion in Eq.~\eqref{expY} - $Y^0$ - corresponds to the limit of unbroken $G_f$. No model can accommodate the observed fermion mass spectrum
in the unbroken phase and several terms in the expansion are needed to provide a realistic description of the data, which typically implies many independent parameters. Depending on the size of $|\tau_\alpha|$, the expansion can converge more or less rapidly. Additional parameters might affect the physical masses, such as those coming from noncanonical kinetic terms
or from RGE effects.
More critically, both the magnitude and the orientation in flavor space of the $\tau_\alpha$ VEVs must be carefully chosen to reproduce the experimental data. Achieving such vacuum alignment is straightforward in the simple U(1)$_{F}$ example discussed earlier, but it becomes increasingly complex in more ambitious models involving elaborate symmetry-breaking sectors. While it is always possible to impose the desired $\tau_\alpha$ VEVs by hand, the more intricate the symmetry-breaking structure becomes, the less transparent and meaningful the overall construction appears.

Many such models have been built, each with intriguingly interesting aspects, but none emerging as more convincing than the others. Most of them employ a large number of independent parameters and require a dedicated symmetry-breaking sector, often characterized by significant complexity. Moreover, they fail to explain why there are exactly three families. The primary utility of this approach perhaps lies in models with a scale of new physics $\Lambda_{NP}$ close to energies accessible today. These models generally predict rare transitions that, in the absence of approximate flavor symmetries, would conflict with experimental data. For example, a high degree of protection against FCNCs is achieved in scenarios with a maximal flavour symmetry, broken only by minimal sources of flavour violation~\cite{DAmbrosio:2002vsn}.
Recent reviews of models of fermion masses based on flavor symmetries can be found in Ref.~\cite{Feruglio:2019ybq,Ding:2024ozt}.
\begin{figure}[h!]
	\centering
	\includegraphics[width=.75\textwidth]{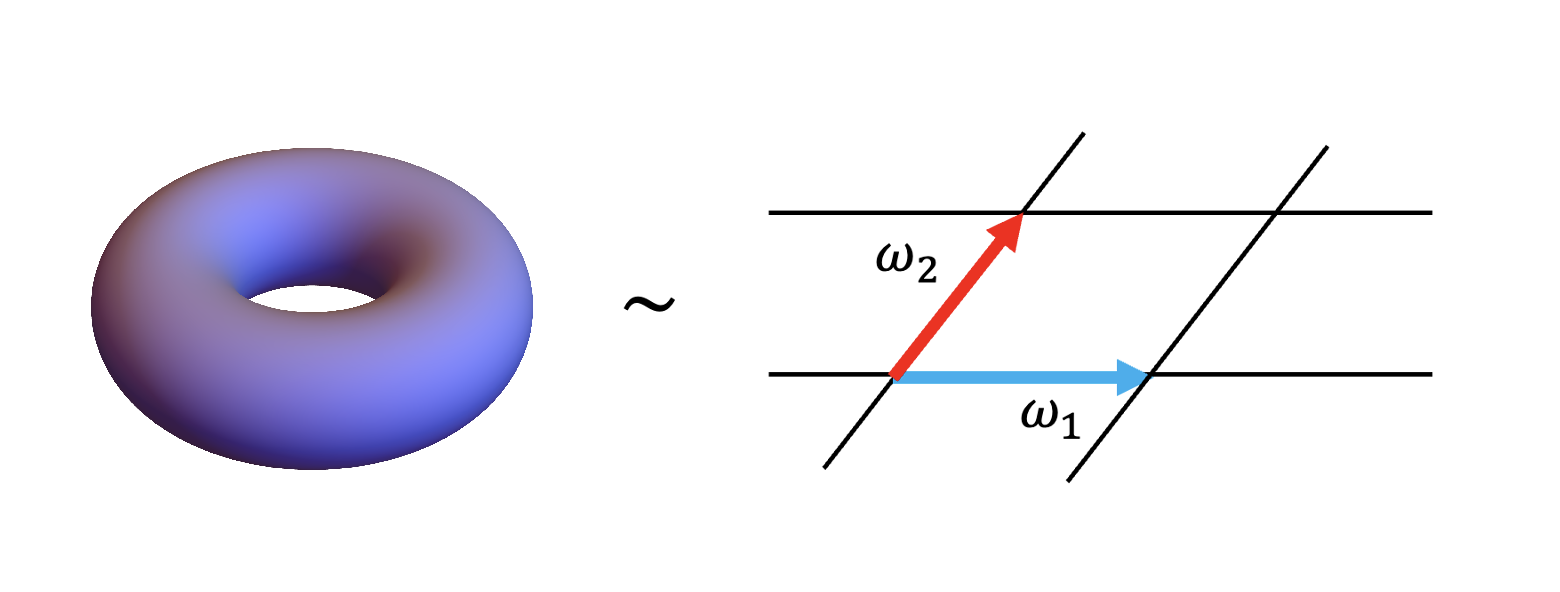}
	\caption{\label{torus} A 2-dimensional torus represented as the complex plane $\mathbbm{C}$ modulo a lattice generated by ($\omega_1,\omega_2)$.}
\end{figure}

Instead of demoting the choice of the symmetry breaking sector to the very last step, one can start the construction by looking for some physically motivated and hopefully uncomplicated such sector. A simple example emerges
adopting as flavor symmetry the modular group $\mathrm{SL}(2,\Z{})$, acting on a complex field $\tau$, $({\tt Im}\tau>0)$ as
\begin{align}
\label{gamma}
\tau\xrightarrow{\gamma}\gamma\tau\equiv\frac{a\tau+b}{c\tau+d}\,,
\end{align}
with  $a,b,c,d$ integers, satisfying $ad-bc=1$. The modulus $\tau$ can be related to the geometry of the torus.
A torus can be described by the complex plane $\mathbbm{C}$, with the identification of points $y$ and $y'$ related by
\begin{align}
\label{equivrel}
y'=y+n_1 \omega_1+n_2 \omega_2\,,
\end{align}
where $\omega_{1,2}$ are complex numbers with ${\tt Im}(\omega_1/\omega_2)>0$ and $n_{1,2}$ are integers, see Fig.~\ref{torus}.
The set $(\omega_1,\omega_2)$ is not uniquely defined, because the choice
\begin{align}
\left(\begin{array}{c}\omega'_1\\\omega'_2\end{array}\right)=
\left(\begin{array}{cc}a&b\\c&d\end{array}\right)
\left(\begin{array}{c}\omega_1\\\omega_2\end{array}\right) \;,
\end{align}
with $a$, $b$, $c$, $d$ integers and $ad-bc=1$, is equally valid to define the equivalence relation of Eq.~\eqref{equivrel}.
Since, by making use of rotations and scale transformations, we can bring any basis $(\omega_1,\omega_2)$ to the form $(\tau,1)$, we find that two tori described by $\tau$ and $\gamma\tau$ are equivalent.
From this viewpoint, the flavor symmetry $\mathrm{SL}(2,\Z{})$ can be thought as a gauge symmetry, removing the redundancy 
related to the use of the modular parameter $\tau$ to describe a torus. The modular group is ubiquitous in string theory,
where moduli are variables describing the size and shape of the compact space.  
Moreover, the nonlinear nature of modular transformations can lead to relations among the coefficients $Y^0$, $Y^\alpha$, $Y^{\bar\alpha}$, $Y^{\alpha\beta}$,... above, potentially increasing predictability.

In a SUSY model, $\tau$ and matter fields $\varphi$ are promoted to chiral supermultiplets,
and the most general transformation of $\varphi$ under $\mathrm{SL}(2,\Z{})$ reads
\begin{align}
\label{eq:ModTrafoFields}
  \varphi\xrightarrow{\gamma}(c\tau+d)^{k_\varphi}\,\rho_\varphi(\gamma)\,\varphi\,,
\end{align}
where $k_\varphi$ is the weight and $\rho_\varphi(\gamma)$ is a unitary representation of a finite copy $\Gamma_N$ of $\mathrm{SL}(2,\Z{})$. 
Asking invariance of the theory under $\mathrm{SL}(2,\Z{})$, one finds that in general the Yukawa couplings are functions of $\tau$ with transformation properties of the type:
\begin{align}
\label{eq:ModTrafoY}
  Y(\gamma\tau)=(c\tau+d)^{k_Y}\rho_Y(\gamma)\,Y(\tau)\,,
\end{align}
with appropriate $k_Y$ and $\rho_Y(\gamma)$. The advantage of a SUSY model is that $Y(\tau)$ only depends on $\tau$, not on $\bar\tau$.
Holomorphic functions with this property are modular forms that,
for given weight $k_Y$ and group $\Gamma_N$, span finite-dimensional linear spaces.
Once $k_Y$ and $\rho_Y(\gamma)$ are chosen, only a few linearly independent sets $Y(\tau)$
exist, and the number of independent parameters in the Yukawa sector can be reduced compared to the FN case.
Moreover, the Yukawa couplings depend on a single symmetry-breaking complex parameter, the VEV of $\tau$,
which can also be responsible for the spontaneous breaking of \CP~\cite{Novichkov:2019sqv,Baur:2019kwi}, thus greatly simplifying the vacuum alignment problem.

This framework~\cite{Feruglio:2017spp} has inspired a wide range of models describing both the lepton and the quark sectors~\cite{Kobayashi:2023zzc,Ding:2023htn}. Yukawa couplings are expressed in terms of a reduced set of continuous parameters, compared to models with linearly realized symmetries.  
Mass hierarchies can be naturally generated when $\tau$ approaches points of enhanced symmetry in the moduli space~\cite{Novichkov:2021evw}.
On the downside, important elements - such as parameters related to the choice of kinetic terms for matter fields~\cite{Chen:2019ewa} - are often neglected. Furthermore, in a purely bottom-up approach, the freedom in choosing $k_\varphi$ and the representation $\rho_\varphi(\gamma)$ is typically too large to clearly identify a minimal or benchmark model. 
String theory, with its restrictive rules, might offer a better scenario for the implementation of these ideas.

\paragraph{Summary}
Broken flavor symmetries represent a useful tool to analyze the pattern of fermion masses and mixing angles.
Models based on abelian symmetries are effective in explaining masses and mixing angles at the order-of-magnitude
level, and are also compatible with GUTs. More elaborate models aiming at a greater predictability, often suffer
from a complicated symmetry breaking sector, associated with a vacuum alignment problem.
Modular invariance, in its simplest version, offers the advantages of an economic symmetry breaking sector
and of a strong motivation rooted in string theory.
In general, in a bottom-up approach, flavor symmetries can be implemented in a large number of different
models, none of which seems to be singled out by data.

\section{Fermion masses in String Theory}\label{sec8}
The origin of quarks and leptons, as well as the explanation of their masses and mixing angles, may lie within the framework of string theory.
In this theory, all elementary particles are modeled as 1-dimensional strings possessing quantum properties. The fundamental energy scale associated with these strings is typically close to the Planck scale, $M_{Pl}$. A striking implication of string theory is that the Universe must contain more than the three familiar spatial dimensions, leading typically to a total of nine spatial dimensions in addition to time. Moreover, in most scenarios, string theory incorporates SUSY. If string theory accurately describes Nature, then the six additional spatial dimensions must have undergone a process of compactification at some stage in the cosmological past. This means that an unknown mechanism would have caused these extra dimensions to build compact geometries of such a small size that they remain undetectable by current experimental endeavors.

Strings are perceived at lower energies as particles, which in the usual quantum terms, correspond to perturbations or quanta of effective fields. That is, each string is associated with a field at low energies. Some of the strings correspond to gauge fields, such as those of the SM or GUTs, while some other strings behave as other matter fields, fermions and bosons. Interestingly, a set of closed strings describes all the degrees of freedom of a graviton, which corresponds to the likely mediator of interactions via quantum gravity. This is the key observation that leads researchers to consider string theory as a promising candidate to describe simultaneously all fundamental interactions, including gravity, and elementary matter.

In addition to strings capable to move in the whole space-time, string theory offers dynamical $p$-dimensional hyperplanes known as D-branes, with $-1\leq p\leq9$. The ends of some open strings, which otherwise are forbidden in the theory, are fixed within these D-branes. Such open strings behave at lower energies as gauge fields, so that stacks of $N$ of these D-branes, all piled within each other, can yield gauge fields associated with a \U{N} gauge symmetry. In some configurations $\mathrm{USp}(2N)$ or \SO{2N} gauge groups can arise too. 

As we shall see in the following subsections, there are remarkable efforts that have led to models reproducing several properties of the SM, including the production of large classes of models endowed with the gauge symmetry of the SM, $G_{SM}$, and three complete generations of quarks and leptons. In all of these constructions, the geometric (and topological) features of the compactification spaces, along with the details of the interactions of strings and branes, deliver a restricted set of discrete symmetries of the 4-dimensional theory that control masses and mixings of the theory. That is, the compactification schemes seem to provide a plausible microscopic origin of flavor symmetries, as introduced in section~\ref{sec7}, reducing somewhat the plethora of admissible flavor groups to those that satisfy stringent conditions associated with string theory. Despite this success, it seems fair to state that these models have not yet brought us closer to the ambitious goal of a unique explanation for the replication of generations and the origin of the patterns of masses and mixings observed in particle physics.

\subsection{Aiming at the SM from strings}

The field known as string phenomenology makes use of the previous ingredients to address the question of how to go from the various string theories to models that describe observed particle physics while providing testable predictions. Depending on their specific properties in 10 dimensions, there are five SUSY string theories: type I with \SO{32} gauge group, type IIA and type IIB (with no embedded gauge groups), and the two heterotic strings with gauge groups \E8\x\E8 and \SO{32}. All of these exhibit $\mathcal N=1$ SUSY (or rather its gravitational extension known as supergravity (SUGRA)), except for the type II theories which have $\mathcal N=2$.

Since we frequently aim at arriving at models in four dimensions with $\mathcal N=1$ SUSY, the geometry of the compact spaces must be that of a so-called Calabi-Yau manifold or a toroidal orbifold. A Calabi-Yau manifold is a 6-dimensional space whose curvature Ricci scalar vanishes everywhere, but exhibit a \SU3 holonomy group. A toroidal orbifold can be understood as a compactification space built in two steps: first, the extra dimensions take the form of a 6-dimensional torus $T^6$ and, then, a discrete set of its isometries building a so-called point group $P$ is modded out, such that $T^6/P$ is the orbifold. This process produces a number of curvature singularities on the orbifold. In type II theories, we eliminate one SUSY by imposing additional constraints known as orientifold, which drops out all strings that do not leave the model invariant when their orientation is reversed.

In the cases of the type I or the heterotic strings, the compactification of string theories on the mentioned spaces does not only reduce the number of observed dimensions, but breaks down the original gauge symmetries. This means that the 4-dimensional SM gauge group $G_{SM}$ can arise as the unbroken remnant subgroup of the original 10-dimensional gauge groups \E8\x\E8 or \SO{32}, once the gauge fields are required to be invariant under the geometric operations associated with the compactification. Beyond the massless strings corresponding to the remnant gauge fields, there are massless and massive strings that build the 4-dimensional matter fields. In particular, depending on the specifics of the compactification scheme, the massless strings and, hence, their related fields may transform under $G_{SM}$ just as quarks and leptons.

In type II theories, as we shall shortly review in section~\ref{sec:Dbranes}, a set of different stacks of D-branes, intersecting at various angles within the compact space and overlapping in our observable Minkowski space-time $\mathds M^4$, opens up a possibility to yield a potentially realistic gauge group. For instance, three separate stacks of D-branes containing three, two and one D-branes, respectively, leads to a global $\U3\x\U2\x\U1\supset G_{SM}$. Moreover, open strings whose ends are attached at two stacks of D-branes are known to transform under the fundamental or anti-fundamental representations of both the gauge groups inherent to the stacks, implying that these strings can be regarded as matter fields at low energies.

It is interesting that the geometry of the 6-dimensional compact space is characterized by moduli, some of which are identical in nature to those introduced in section~\ref{sec:variants}. That is, properties such as the size and shape of the space used for the compactification depend on the VEVs of the available moduli, whose transformations build modular groups that are symmetries of the compact space and thus of the emerging 4-dimensional effective model. As we shall briefly explore in section~\ref{sec:Eclectic}, it is also possible that the geometric structure of the compact space yields some purely geometric global transformations of the matter fields. These properties may hint towards a plausible origin of flavor symmetries in string compactifications. We now study some of the details that let us arrive at masses of quarks and leptons in some compactification schemes.

\subsection{Eclectic flavor symmetries from heterotic orbifolds}
\label{sec:Eclectic}

\begin{figure}[b!]
    \centering
    \begin{footnotesize}
       \includegraphics{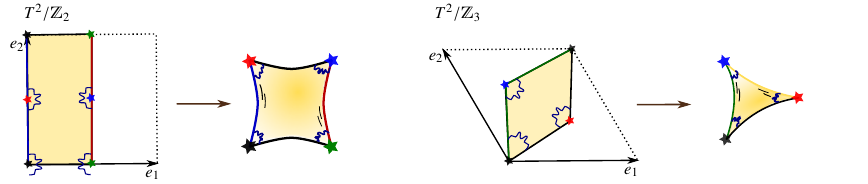}
    \end{footnotesize}
    \caption{\label{fig:orbifolds} Strings attached to singularities of 2-dimensional toroidal orbifolds. $T^2$ is generated by the lattice vectors $e_i$. The singularities are fixed points under the combined action of the abelian point groups \Z2 (left) and \Z3 (right) and lattice translations. The strings close due to the orbifold action.}
\end{figure}

Heterotic orbifolds~\cite{Dixon:1985jw,Dixon:1986jc} provide a fertile landscape for flavor phenomenology.
In these constructions, the extra dimensions of the heterotic string are supposed to adopt the compact shape of a 6-dimensional toroidal orbifold $T^6/P$ and a very small fundamental size scale, very close to  the Planck length, $\ell_{Pl}\approx1.6\times10^{-35}$\,m. These spaces are flat everywhere, except for a few special points known as singularities, which concentrate all of their curvature and are left fixed by the action of $P$. For example, consider a 2-dimensional subsector $T^2$ modded out by an abelian \Z2 or \Z3 point group; after the point group action on the torus, the orbifold takes the shape of a pillow-like object with four or three curvature singularities, see Fig.~\ref{fig:orbifolds}. Note that $T^6/P$ can equivalently be expressed as $\mathds R^6/S$, where the so-called {\it space group} $S$ of the orbifold combines the action of the point group $P$ and the lattice shifts $\{e_i\}, i=1,\ldots,6$, associated with the lattice basis of the torus. All admissible space groups $S$ leading to consistent 4-dimensional SUSY effective theories have been classified~\cite{Fischer:2012qj}.

Matter on heterotic orbifolds arises from closed strings, as no open strings are admitted.
Before compactification on an orbifold, these include only the gravity supermultiplet and the \E8\x\E8 or \SO{32} super-Yang Mills gauge multiplets.
After compactification on an orbifold some aspects change. First, beyond the dimensional reduction to four dimensions,
only a subset of states corresponding to gauge bosons are left invariant, yielding a 4-dimensional gauge group $G_{4D}\subset\E8\x\E8$ or \SO{32}. Secondly, a new set of closed strings arises. In terms of the lattice description of a torus of Fig.~\ref{torus}, in our illustration of a 2-dimensional orbifold in Fig.~\ref{fig:orbifolds} we see at the singularities that there are some strings that appear to be open in the torus, whose ends are nevertheless identified after performing a \Z2 (180-degree) or \Z3 (120-degree) rotation and applying some torus translations by integer multiples of the toroidal lattice generators $e_i$. This is obvious e.g.\ at the origin (black star) of the \Z2 orbifold lattice. These so-called {\it twisted} strings are matter states that are free to move in the uncompactified space-time $\mathds M^4$, but are fixed in the orbifoldized dimensions. 
Thirdly, besides the gauge bosons, some of the untwisted states that are left invariant by the space-group action become complex fields called moduli, associated with the shape and size of the compact space. For each 2-dimensional subsector, there is one (complex structure) modulus $U$ characterizing the shape of the orbifold,
and one K\"ahler modulus $T$ that set the size of the compact space. While $U$ must develop a VEV $\langle U\rangle$ for all \Z{K} point groups with $K>2$ to be consistent with the orbifold~\cite{Nilles:2020gvu}, $T$ is unfixed and admits all modular transformations in $\mathrm{SL}(2,\Z{})$.

Interestingly, by carefully choosing $S$ and its embedding into the gauge degrees of freedom, one can find many models where $G_{4D}=G_{SM}$ and twisted matter states transform as quarks and leptons, see Table~\ref{tab:SM}. It seems convenient to choose a model with a space group leading to a $T^2/\Z3$ sector endowed with three singularities, as such a geometry would naturally explain the existence of exactly three SM generations.

In this context, the theory provides a set of constraints on the interactions that define flavor symmetries.
As we have seen, a $T^2/\Z3$ orbifold can take the shape of an equilateral triangle. The symmetries of this geometry build the group $S_3$. It is interesting that the process of  abelianization~\cite{Ramos-Sanchez:2018edc} of the corresponding space group leads in addition to a \Z3\x\Z3 discrete symmetry, which enhances $S_3$ to $\Delta(54)\cong S_3\ltimes(\Z3\x\Z3)$, under which the strings localized at the orbifold singularities build representations. This can be generalized to all orbifold space groups~\cite{Kobayashi:2006wq,Nilles:2012cy,Olguin-Trejo:2018wpw}. These symmetry groups do not depend on the size and shape of the compact space, i.e.\ they are moduli independent. We call these symmetries traditional flavor symmetries.

A crucial observation is that these symmetries are obeyed by the couplings $\Phi_1\Phi_2\Phi_3\ldots$ among 4-dimensional effective superfields $\Phi_a$, whose magnitudes can in principle be exactly computed via CFT correlation functions mediated by worldsheet instantons~\cite{Dixon:1986qv}. These coupling strengths of twisted fields turn out to be modular forms that depend on the free moduli and transform non-trivially under modular transformations~\cite{Lauer:1989ax,Lauer:1990tm}. Unfortunately, computing the coupling strengths in general is not easy, which explains why they are known only at the lowest orders~\cite{Hamidi:1986vh}. One must find a more direct method to identify the couplings and their transformation properties.

An alternative to arrive at the couplings and the structure of the action is to consider the outer automorphisms of the space group $S$ in the Narain formalism~\cite{Narain:1985jj,GrootNibbelink:2017usl}. In this formalism, spatial coordinates are split into left and right independent components as the structure of heterotic strings is based on the fact that, for closed strings, right and left-moving modes are independent. This doubles the effective coordinates implying that elements of the space group $S$ in a 2-dimensional orbifold are now expressed as 4-dimensional rotational and translational elements. An important observation is that the group of automorphisms of $S$ are symmetries of the orbifold. Clearly, only outer automorphisms represent symmetries that are not introduced in $S$ itself. It is remarkable that translational outer automorphisms coincide with traditional flavor symmetries, leaving the metric of the compact space untouched (as expected). Simultaneously, rotational outer automorphisms define the modular transformations associated with the K\"ahler and complex structure moduli of the theory. It is precisely the doubling of coordinates associated with the Narain formalism what explains these two types of modular symmetries. In a $T^2/P$ sector, they lead to the $\mathrm{SL}(2,\Z{})$ symmetry of the toroidal $T$ modulus as well as the unbroken modular transformations of $U$ due to the orbifold fixing of its VEV. Notably, twisted matter states located at different singularities mix under these modular symmetries as in Eq.~\eqref{eq:ModTrafoFields}~\cite{Lauer:1989ax,Lauer:1990tm}, building representations of $\Gamma_N'$ (double cover of $\Gamma_N$) finite modular groups. This identification allows one to determine all modular forms of the theory: one must simply consider the modular forms associated with $\Gamma_N'$ and their transformations~\eqref{eq:ModTrafoY}. The combination of both traditional and modular symmetries, appearing naturally in heterotic orbifolds~\cite{Baur:2019iai}, is called {\it eclectic} flavor symmetry.

Eclectic flavor symmetries combine the power of both traditional and modular symmetries. While the modular symmetries reduce the number of parameters in Yukawa couplings, the traditional flavor symmetries remove the ambiguities associated with the otherwise unavoidable great freedom left by modular symmetries in the kinetic terms~\cite{Chen:2019ewa}. The price to pay is that annoying flavon fields must be introduced to break the traditional sector of the eclectic group (which leads to predictions), reintroducing the issues associated with the choice of the VEVs of such fields.

For concreteness, let us briefly discuss the case of the sector $T^2/\Z3$. The method based on the orbifold outer automorphisms leads to a $\Gamma_3'\cong T'$ modular flavor symmetry, which, combined with the $\Delta(54)$, builds the eclectic flavor symmetry $\Omega(1)$ of order 648. The VEV of the complex structure modulus is $\langle U\rangle=\omega=e^{2\pi\I/3}$ leaving an unbroken \Z3 symmetry in the sector, while the K\"ahler modulus $T$ admits all $\mathrm{SL}(2,\Z{})$ transformations. All properties of the states and the modular forms are fully determined by the compactification. The three twisted matter fields have fractional modular weight $-2/3$, and transform as a $\rep{3}_2$ under $\Delta(54)$ and as $\rep2'\oplus\rep1$ of $T'$. (The fractional modular weights imply that the unbroken \Z3 related to $U$ transformations is actually realized as a $\Z9^R$ $R$-symmetry by matter fields.) Further, the lowest-weight modular forms depend only on the $T$ modulus and build a $\rep2''$ of $T'$ with weight 1. This illustrates that string theory imposes intriguing constraints that are not available in bottom-up constructions.
Phenomenology with this particular scenario has been explored explicitly in one particular string-derived model with three SM matter generations~\cite{Baur:2022hma}. It still remains to be seen how general the emerging predictions are and how they can be tested experimentally. Also, the challenging issue of stabilizing the moduli and flavon VEVs remains to be done.

\subsection{Flavor in models with D-branes}
\label{sec:Dbranes}

\begin{figure}[t!]
    \centering
    \begin{footnotesize}
       \includegraphics{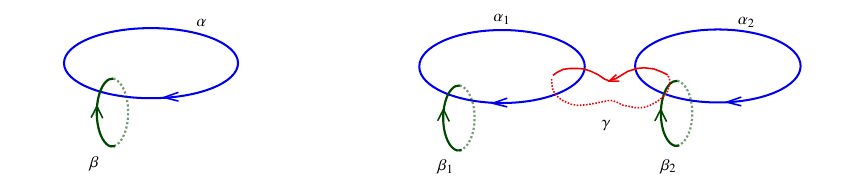}
    \end{footnotesize}
    \caption{\label{fig:tori} The various 1-cycles in a 2-dimensional torus and a 2-holed torus.}
\end{figure}

Let us now explain flavor symmetries in the context of type II string theories with D-branes. We start by inspecting the origin of a semi-realistic 4-dimensional spectrum, which includes the SM gauge group together with states transforming as generations of quarks and leptons (see e.g.~\cite{Ibanez:2001nd,Blumenhagen:2005mu}). In this case, the compactification scheme consists of a number of steps. First, as described before, the six extra dimensions are compactified on either an orbifold or a Calabi-Yau, imposing in addition an orientifold in order to obtain $\mathcal N=1$ SUSY in four dimensions. To achieve the SM gauge group and matter, one typically considers a number of stacks of $(3+q)$-dimensional D-branes, $1\leq q\leq6$, which overlap in the three observable spatial dimensions of our Minkowski space-time $\mathds{M}^4$ and wrap non-trivial $q$-cycles of the compactification space. $q$-cycles may be graphically understood as closed $q$-dimensional paths around handles and holes of the compact space; e.g.~a 2-dimensional torus and a 2-holed torus have the 1-cycles illustrated in Fig.~\ref{fig:tori}.

\begin{figure}[b!]
    \centering
    \begin{footnotesize}
       \includegraphics{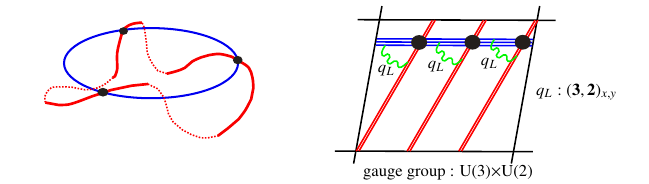}
    \end{footnotesize}
    \caption{\label{fig:DbraneExample} Two D-brane stacks intersecting at angles in 1-cycles of a torus of the compact space to produce three quark doublets $q_L:(\rep3,\rep2)_{x,y}$ (up to an extra \U1 charge and proper normalization) at the intersections of the D-branes.}
\end{figure}

While overlapping in $\mathds M^4$ where observable physics takes place, the stacks of $p$-dimensional D-branes may wrap different $q=(p-3)$-cycles in various directions. To understand how this leads to matter fields and their replication, consider again a 2-dimensional sector in the compact space of a model based on a 6-dimensional (factorizable) torus, $T^6=T^2\x T^2\x T^2$. Using the lattice description of Fig.~\ref{torus}, we study a situation in which a stack of three D-branes wraps a 1-cycle of a torus while a stack of two D-branes wraps another 1-cycle of the same torus at an angle w.r.t.\ the first 1-cycle, as depicted in Fig.~\ref{fig:DbraneExample}. Besides the open strings that build a $\U3\x\U2\cong\SU3\x\SU2\x\U1^2$ gauge group, the open strings whose ends are attached to both stacks, arising at the intersections, transform as the bifundamental gauge representations $(\rep3,\rep2)_{x,y}$, where $x,y$ denote here some \U1 charges. One can easily find a basis in which $x$ and $y$ combine to give the hypercharge of a quark doublet $q_L$, see Table~\ref{tab:SM}. The angle between the D-brane stacks is chosen such that, using the properties of the torus, we find a triple intersection and, hence, $q_L$ is replicated three times as required by observations. In general, any $p$-dimensional D-brane stack $\alpha$ wraps a $q$-cycle $\Pi_\alpha$ in the compact space, and, hence, can intersect several times with some other D-brane stack $\beta$ wrapping another $q$-cycle $\Pi_\beta$. The multiplicity of intersections provides the net number of chiral matter states. This is given by the topological invariant intersection number $I_{\alpha\beta}:=[\Pi_\alpha]\cdot[\Pi_\beta]$ in terms of the homology classes $[\Pi_i]$ of the associated $q$-cycles.

\begin{figure}[t!]
    \centering
    \begin{footnotesize}
       \includegraphics{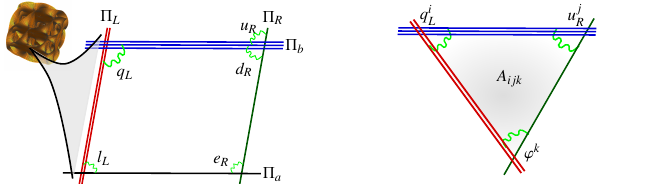}
    \end{footnotesize}
    \caption{\label{fig:fullDbraneExample} Zoom in of the chiral matter spectrum (left) located at some 2-dimensional subspace of a Calabi-Yau manifold, where four stacks of 6-dimensional D-branes intersect providing one SM generation in a type IIA string model. At some other 2-dimensional subspace (right), a different intersection structure yields an up-quark Yukawa coupling, which is proportional to $e^{-A_{ijk}}$, where $A_{ijk}$ is the moduli-dependent area of intersection.}
\end{figure}

There are examples of more realistic scenarios with three full SM generations. They are based on type IIA strings with either Calabi-Yau or orbifold (orientifold) compact space and four stacks of $(3+3)$-dimensional D-branes wrapping different 3-cycles and intersecting as illustrated in Fig.~\ref{fig:fullDbraneExample} (left). Although only one SM family is shown,
the net number of three SM families is obtained by carefully choosing the various cycles, so that all intersection numbers satisfy $|I_{\alpha\beta}|=3$, see e.g.~\cite[Section 4]{Cremades:2003qj}. 

Yukawa couplings in these scenarios result of open-string worldsheet instantons connecting three D-brane intersections~\cite{Cremades:2003qj}, somewhat similar to couplings in heterotic orbifolds. A situation leading to an admissible Yukawa couplings is illustrated in Fig.~\ref{fig:fullDbraneExample}: there is a small locality with area $A_{ijk}$ in the compact space, where fields with quantum numbers combining to build a gauge invariant coupling appear at three intersections. In fact, Yukawa couplings turn out to depend on the K\"ahler moduli and not on the complex structure, as happens in most cases of heterotic orbifold models. Also like in heterotic orbifolds, couplings may be expressed as modular functions: they adopt the shape of Jacobi $\vartheta$-functions.
It is possible to conceive a series of appropriate $q$-cycles, such that all phenomenologically required couplings are available.

These models are naturally embedded with discrete non-abelian symmetries~\cite{Marchesano:2013ega}.
One finds very frequently that the traditional flavor symmetry $D_4$ or multiple products of this group constrain the masses of quarks and leptons in this kind of models. This result is not very surprising since, as mentioned before, one can start with an orbifold and then include D-branes and orientifolds to arrive at viable models. In fact, one common arrangement includes a sector with a $T^2/\Z2$, which leads to the geometric (traditional) flavor symmetry $D_4$, see Fig.~\ref{fig:orbifolds} left. There is yet a caveat. As we shall shortly see, in compactifications on magnetized tori, the flavor symmetry is modular and leads particularly to metaplectic flavor groups. As discussed in e.g.~\cite{Kobayashi:2016ovu}, the structure of the Yukawa couplings in magnetized toroidal compactifications can be dualized to give us the couplings in intersecting D-brane models. So, it is to expect that, on top of the mentioned traditional symmetry, some modular flavor symmetry may appear in this case. However, some work is still needed to provide the full flavor structure of this scenario. A natural question is whether its properties would resemble those of eclectic flavor groups.

\subsection{Metaplectic flavor symmetries from magnetized branes}
\label{sec:Metaplectic}

Another interesting top-down scenario, where non-abelian flavor symmetries naturally arise is based on compactifications of type IIB strings on tori with background magnetic fluxes. Although these models are considered to be dual to some intersecting D-brane scenarios, computing the Yukawa couplings and hence the masses of SM matter fields arising in these models turns out to provide a perhaps clearer picture, which reveals the modular properties of the theory. The study of the details of these constructions was pioneered in~\cite{Cremades:2004wa}.

Let us consider, as we have done so far, only a 2-dimensional sector of the compact space of a model with a \U{N} gauge symmetry. The extra dimensions are compactified on a torus endowed with a magnetic flux, which is associated with the coupling between the charges and the gauge bosons. We assume that the flux leads to the breakdown of the gauge symmetry $\U{N}\to\U{N_a}\x\U{N_b}\x\U{N_c}$, with $N=N_a+N_b+N_c$. In~\cite{Cremades:2004wa}, it was shown that magnetic fluxes induce massless left-chiral fermions that can be described by the wavefunctions
\begin{equation}
 \label{eq:ZeroModesWavefunctions}
 \wavefunction(z,\tau,\zeta)~=~ \mathcal{N}\,
 e^{\pi\,\I\,M\,(z+\zeta)\,\frac{\im (z+\zeta)}{\im\tau}}
 \,\vartheta\orb{\frac{j}{M}}{0}\bigl(M\,(z+\zeta),M\,\tau\bigr)\,,
\end{equation}
where the integer $M$ counts the units of flux carried by the field, $0\le j\le M-1$ is an integer. Further, $z$ denotes the coordinate in the extra dimensions of the torus, $\zeta$ is a so-called Wilson line parameter, $\tau$ denotes the modulus of the torus, $\vartheta$ denotes the Jacobi $\vartheta$-function, and $\mathcal N$ is a modular-dependent normalization factor. The wavefunctions~\eqref{eq:ZeroModesWavefunctions} living in the compact space are associated to 4-dimensional matter superfields $\Phi^{j,M}$.

It was also shown that the 4-dimensional Yukawa-coupling strength of the three massless superfields associated with the wavefunctions, $\Phi^{i,\mathcal{I}_{ab}}\Phi^{j,\mathcal{I}_{ca}}\Phi^{k,\mathcal{I}_{cb}}$, is given in the absence of Wilson lines ($\zeta=0$) by~\cite{Almumin:2021fbk}
\begin{equation}
\label{eq:Yijk}
 Y_{ijk}(\tau)~=~-\,\int\limits_{T^2}\!d^2 z\,
 \wavefunction[i,\mathcal{I}_{ab}](z,\tau,0)\,
 \wavefunction[j,\mathcal{I}_{ca}](z,\tau,0)\,
 \left(\wavefunction[k,\mathcal{I}_{cb}](z,\tau,0)\right)^* ~\propto~ \vartheta\orb{\hat\alpha_{ijk}/\lambda}{0} (0,\lambda\tau)\,,
\end{equation}
where $\wavefunction[i,\mathcal{I}_{ab}](z,\tau,0)$ transforms in the bifundamental representation $(\rep{N_a},\crep{N_b})$ of $\U{N_a}\x\U{N_b}$, the multiplicities $\mathcal{I}_{ab}$ are some integers such that $\mathcal{I}_{ab}+\mathcal{I}_{bc}+\mathcal{I}_{ca}=0$, $\lambda = \mathrm{lcm}(|\mathcal{I}_{ab}|,|\mathcal{I}_{bc}|,|\mathcal{I}_{ca}|)$ and $\hat\alpha_{ijk}$ is some integer that depends on the multiplicities $\mathcal{I}$ and the counters $i,j,k$. The relevant feature of $Y_{ijk}(\tau)$ is that it transforms as a modular form with fractional weight $1/2$. This implies that the group of transformations governing modular forms builds the metaplectic group $\tilde\Gamma=\mathrm{Mp}(2,\Z{})$, i.e.\ the double cover of $\Gamma=\mathrm{SL}(2,\Z{})$. Consequently, they are components of vector-valued modular forms of metaplectic finite modular symmetries (of level $4N$, $N\in\mathds{N}$),
\begin{equation}
\label{eq:ModularTransformationYukawas}
 Y^{\hat{\alpha}}_{ijk}(\tau) ~\xmapsto{~\tilde\gamma~}~ Y^{\hat{\alpha}}_{ijk}(\tilde\gamma\,\tau)
  ~=~ \pm(c\,\tau+d)^{1/2}\,\rho_{\rep{\lambda}}(\tilde\gamma)_{\hat{\alpha}\hat{\beta}}\,Y^{\hat{\beta}}_{ijk}(\tau)\,.
  \qquad \tilde\gamma\in\tilde\Gamma\,,
\end{equation}
Note that the automorphy factor differs from the standard $\mathrm{SL}(2,\Z{})$ transformation~\eqref{eq:ModTrafoY} by the inclusion of extra signs.
On the other hand, explicitly computing the modular transformations of the 2-dimensional wavefunctions shows that they exhibit modular weight $1/2$. Since $\psi\otimes\Phi$ must have trivial modular properties, matter superfields $\Phi^{j,M}$ transform similar to Yukawa couplings with weight $-1/2$,
\begin{equation}\label{eq:ModularTransformation4DFields_1}
  \Phi^{j,M} ~\xmapsto{\tilde\gamma~}~ \pm(c\,\tau+d)^{-1/2}\,\rho_{M}^\phi(\tilde\gamma)_{jk}\,\Phi^{k,M}\;.
\end{equation}
Once more, we see that the properties of matter fields and Yukawa couplings are determined by the characteristics of the compactification.

It is possible to conceive some configuration in this scheme that may lead to a semi-realistic spectrum of the model. More progress in this field must be done to uncover the predictions and details that it can offer to the search of a solution for the origin of masses of quarks and leptons.

\paragraph{Summary}

String theory offers different mechanisms to generate models with both the gauge group and the matter spectrum of the SM. In addition, the properties of the compactification delivers various possibilities to arrive at non-abelian discrete flavor symmetries. Interestingly, (moduli-independent) traditional flavor symmetries and modular symmetries appear naturally as a result of the compactification of the extra dimensions. There are scenarios in which both of them have a common origin and offer a combined scheme, known as eclectic flavor group. In all cases, the restrictions on matter fields imposed by internal consistency conditions of string theory translate to fixing the properties of matter fields, such as flavor (traditional and modular) representations and modular weights. In the case of modular symmetries, also modular forms are set by the chosen compactification. It remains as a challenge to uncover whether these additional constraints are enough to deliver better explanations than bottom-up constructions.
\section{Conclusions}
\label{sec:conclusions}

Quarks and leptons are organized into three families that share identical gauge charges but differ in their precisely measured masses, mixing parameters, and \CP-violating effects. This raises a fundamental question: do these patterns point to a deeper organizing principle, or does the SM already represent the most streamlined and accurate framework? Although the latter cannot be ruled out in principle, this chapter has explored efforts to uncover a guiding mechanism that might explain the structure of fermion masses and mixings quantitatively. We have reviewed several scenarios.
GUTs aim to unify quarks and leptons within shared group structures, suggesting relations among their mass-generating interactions. In particular, \SO{10}
frameworks naturally accommodate heavy right-handed neutrinos, offering a plausible account for the lightness of neutrino masses through the seesaw mechanism. However, basic GUT models often lack the flexibility to fully match observed fermion masses without introducing many parameters, which weakens their predictive strength. 
One approach to explaining fermion-mass hierarchies is to assume lighter masses arise through radiative corrections, necessitating approximate chiral symmetries that are broken by other sectors. This reframes the flavor problem as one of identifying the origin and properties of the symmetry-breaking dynamics. Models with composite fermions --- bound states in strong dynamics --- can generate small masses naturally, but realistic implementations are scarce, and the required energy scales exceed 10 TeV. A compromise envisions SM fermions as partly composite, with their flavor structure set by mixing with composite partners. This idea, supported by 5D warped-space constructions, captures mass hierarchies geometrically but struggles with flavor-changing constraints. Flavor symmetries provide another avenue for organizing fermion patterns. Simple abelian models can reproduce observed trends and fit within GUTs, while more predictive frameworks often face challenges like complex vacuum structures. Modular symmetries offer a minimal and string-theoretically motivated alternative.
In string theory, the compactification of extra dimensions can naturally produce both the SM spectrum and discrete flavor symmetries. Traditional and modular flavor groups may even share a unified origin, forming so-called eclectic symmetries. These built-in constraints on particle properties raise the possibility of more constrained flavor predictions, though it remains unclear if they outperform phenomenological bottom-up models. We do not know whether one or more of these elements will eventually be incorporated into a fundamental theory, or whether Nature will surprise us by opening up entirely unexpected directions.


\begin{ack}[Acknowledgments]%
We thank Gui-Jun Ding, Tony Gherghetta, Hitoshi Murayama, Jo\~ao Penedo and Arsenii Titov for useful comments.
 The work of FF was supported by INFN.
 The work of SRS was partly supported by UNAM-PAPIIT grant IN113223 and Marcos Moshinsky Foundation.
\end{ack}


\bibliographystyle{Numbered-Style} 
\bibliography{Quarkandleptonmasses}

\end{document}